# Assessing the Climate and Habitability of Tidally Locked Rocky Exoplanets

**Erica Bisesi[1,2,*], Giuseppe Murante[1,2,3,4], Jost von Hardenberg[5,6], José A. Caballero[7], Michele Maris[1,3,4], Daniela Billi[8], Nicoletta La Rocca[9,1], and Laura Silva[1,3]**

[1] INAF – Osservatorio Astronomico di Trieste, Via G. Tiepolo 11, 34143 Trieste, Italy; [2] CNR – Istituto di Geoscienze e Georisorse, Via G. Moruzzi 1, 56124 Pisa, Italy; [3] Institute for Fundamental Physics of the Universe, Via Beirut 2, 34151 Trieste, Italy; [4] Centro di Ricerca Nazionale in High Performance Computing, Big Data e Quantum Computing (ICSC), Via Magnanelli 2, 40033 Casalecchio di Reno (Bologna), Italy; [5] Politecnico di Torino – DIATI, Corso Duca degli Abruzzi 24, 10129, Torino, Italy; [6] CNR – Istituto di Scienze dell'Atmosfera e del Clima, Corso Fiume 4, 10133, Torino, Italy; [7] CSIC-INTA – Centro de Astrobiología, Camino Bajo del Castillo s/n, Campus ESAC, 28692 Villanueva de la Cañada, Torrejón de Ardoz (Madrid), Spain; [8] Università degli Studi di Roma "Tor Vergata", Dipartimento di Biologia, Via della Ricerca Scientifica 1, 00133 Roma, Italy; [9] Università degli Studi di Padova, Dipartimento di Biologia, Via U. Bassi 58b, 35131 Padova, Italy

* Corresponding author: erica.bisesi@inaf.it; tel.: +39 040 3199233

## Abstract

Tidally locked exoplanets orbiting M dwarf stars are prime targets in the search for habitable worlds, yet their complex climate dynamics challenge conventional models of habitability. We homogeneously estimated the liquid-water habitability of exoplanets with the highest up-to-date Earth-similarity index. We modified the climate model `PLASIM` to apply it to tidally locked exoplanets across a range of parameters. We systematically classified and compared potentially habitable exoplanets and selected among the ones featuring the highest Earth-similarity index in the conservative sample of the Habitable Wolds Catalog, both relative to each other and as a function of atmospheric pressure. Our analysis revealed three main climate regimes: Snowball, Hot, and Eyeball planets—the latter having a warm and habitable substellar region with $T > 273$ K and a localized hydrological cycle. Although surface temperature generally increases with pressure, Eyeball planet climates remain stable in a wide pressure range. Complex atmospheric dynamics rules the behavior of the climate, highlighting the need for detailed climate simulations to guide future observations.

## 1. Introduction

### 1.1. Astrobiological context: Habitability and climate of tidally locked exoplanets

The search for habitable worlds is one of the most compelling pursuits in science and is a central theme in astrobiology. Since the discovery of the first exoplanet orbiting a Sun-like star at the end of the 20th century (Mayor and Queloz, 1995), thousands of exoplanets have been identified. These planets exhibit a wide diversity in mass, radius, orbital parameters, and the spectral types of their host stars. Particular attention has been devoted to planets orbiting M-type stars, as these hosts are the most numerous in the galaxy (Bochanski et al., 2010; Reylé et al., 2021) and are especially favorable for the detection of small Earth-sized planets. The ongoing search for habitable worlds focuses on rocky or water-rich planets, especially around late-type stars (Endl et al., 2006; Nutzman and Charbonneau, 2008; Bonfils et al., 2013; Zechmeister et al., 2009; Ricker et al., 2015; Ribas et al., 2023).

In broad terms, planetary habitability emerges from a plethora of evolutionary and contingent geological, atmospheric, and stellar properties (e.g., Spohn et al., 2026). Instead, from a pragmatic astrobiological perspective, a planet is considered potentially habitable if it can sustain liquid water on its surface, a condition primarily governed by surface temperature (Dole, 1964; Kasting et al., 1993; Kopparapu et al., 2013; Barnes et al., 2015; Del Genio et al., 2019; Woodward et al., 2025). This in turn is determined by radiative balance, atmospheric dynamics, the presence and properties of oceans, and topography, as well as climate feedback mechanisms. Assessing exoplanetary habitability therefore requires climate modeling, which explores the effects of stellar, orbital, and planetary factors, especially atmospheric pressure and composition, or surface conditions (e.g., Shields, 2019).

Most known rocky exoplanets in the habitable zones of M dwarf stars are likely tidally locked, with permanent day and night sides. This is because the level of stellar flux needed to sustain surface liquid water places these planets close to their host stars, where tidal forces rapidly synchronize their rotation and revolution, typically resulting in 1:1 spin-orbit resonance and negligible obliquity (Joshi et al., 1997; Tarter et al., 2007). In such configurations, most stellar radiation is received at the substellar point, that is, the location where the star is directly overhead (e.g., Shields et al., 2016; Barnes, 2017).

The key factors that govern heat transport and climate, and therefore surface temperature, on any planet include the angular distance from the substellar point and the planet's rotation and orbital periods, which control the Coriolis force (e.g., Showman et al., 2014; Kaspi and Showman 2015; Noda et al., 2017; Haqq-Misra et al., 2018; Hammond and Lewis, 2021). In fact, previous work has shown that synchronous rotation can substantially alter climate relative to Earth-like rotation. Slowly rotating tidally locked planets may develop a strong day-night overturning circulation with strong ascent and optically thick clouds near the substellar point, which increases planetary albedo and can cool the climate, shifting the inner edge of the habitable zone inward (Yang et al., 2013; 2014; 2023). More broadly, the rotation rate determines the Coriolis parameter and therefore the structure of large-scale overturning and eddy activity, with consequences for the distribution of humidity and precipitation (Del Genio and Suozzo, 1987; Way et al., 2018). Haqq-Misra et al. (2018) classified tidally locked temperate planets into different dynamical regimes (slow rotators, Rhines rotators, and rapid rotators). Because later-type M dwarfs typically host habitable-zone planets on shorter-period orbits, synchronously rotating planets around them are more likely to occupy the rapid rotator regimes. These regimes modulate the structure and efficiency of day-night transport and therefore can shift the stability of the climate of partially glaciated Eyeball-like planets, where the "Eye" is defined as the ice-free substellar region. Atmospheric circulation contributes to shape the climate and planetary habitability together with atmospheric composition, cloud formation and distribution, continental and ocean coverage. Furthermore, since cloud coverage is a dynamic process that directly modulates global thermal emission (Yang et al., 2013; Sergeev et al., 2020), any transition in a planet's dynamical regime inevitably drives surface climate variations via cloud-radiative feedbacks.

### *1.2. Climate modeling of tidally locked planets*

Global Circulation Model (hereafter, GCM) studies further indicate that tidally locked planets can maintain ice-free nightsides when stellar flux, greenhouse forcing, or atmospheric mass are sufficiently high (e.g., Hu and Yang, 2014; Zhang and Yang, 2020; Paradise et al., 2022). Additionally, the reddened spectrum of M dwarfs modulates radiative feedbacks in ways that differ significantly from Earth-like conditions, primarily due to near-infrared atmospheric absorption and lower ice and snow albedos (Paradise et al., 2021; 2022). Given these complexities, three-dimensional GCMs are essential for simulating such climates and have demonstrated that surface habitability is possible under a wide range of conditions (see, e.g., reviews by Shields et al., 2016; Barnes, 2017; Haqq-Misra et al., 2018; Del Genio et al., 2019; Lobo and Shields, 2024).

As shown by Pierrehumbert (2011), synchronously rotating planets may occupy at least two distinct climate states: a globally frozen, uninhabitable state and an Eyeball state, in which only an approximately circular region around the substellar point remains above freezing. Lobo and Shields (2024) extended this picture using `ExoCAM` simulations of Earth-like planets in synchronous, relatively slow rotation around K and M dwarfs, considering both aquaplanets and dry land planets over a range of stellar fluxes within the habitable zone. They identified two habitable regimes: the classical Eyeball configuration and a "terminator-habitable" regime, in which excessive substellar heating confines clement surface conditions to the day-night boundary. This latter regime occurs only for land planets, where the absence of oceans limits surface heat transport, allowing strong day-side heating. Atmospheric pressure and mass provide additional critical controls on these climate states, modulating heat transport, humidity, and atmospheric cooling via Rayleigh scattering, as well as enhancing greenhouse gas absorption through pressure broadening. Consequently, variations in atmospheric mass significantly influence the transition from Eyeball climates to globally deglaciated states, as well as the overall stability of habitable regimes; for example, Macdonald et al. (2025) demonstrated that this occurs because atmospheric mass modulates both heat transport and humidity.

Despite the significant advances in characterizing these climate states, assessing the habitability of such planets remains challenging due to the complexity of their climate systems and the multitude of interdependent processes involved. Efforts such as the TRAPPIST-1 Habitable Atmosphere Intercomparison (`THAI`) project (Fauchez et al., 2020; 2022; Turbet et al., 2022; Sergeev et al., 2022) have addressed these issues by comparing four 3D GCM (`ExoCAM`, `LMD Generic`, `ROCKE - 3D`, `Unified Model`) applied to the climate of TRAPPIST-1 e. Although all models broadly agree in predicting a temperate climate under specific atmospheric assumptions, they show significant discrepancies in key quantities such as surface temperature, cloud distribution, and water vapor content. As emphasized by these studies, such differences highlight the importance of employing a hierarchy of physically grounded climate models, depending on the specific research objective, to balance physical realism and computational efficiency.

Within this hierarchy, intermediate-complexity models (`EMICs`) are particularly valuable for systematic explorations of large parameter spaces. In this context, Paradise et al. (2022) modified and adapted `PLASIM` (`PLanet SIMulator`; Fraedrich et al., 2005; Fraedrich, 2012), an intermediate complexity Earth system model originally developed for Earth's climate, to simulate tidally locked exoplanets orbiting M dwarf stars. The resulting model, `ExoPLASIM`, was validated through comparisons with the `THAI` simulations of TRAPPIST-1 e, showing that it reproduces habitability regimes consistent with the range produced by more complex GCMs. Paradise et al. (2021, 2022) investigated the effect of varying the partial pressure of $N_2$ (0.1–10 bar) while keeping the $CO_2$ amount constant. They found that atmospheric pressure significantly influences climate by enhancing greenhouse gas absorption through pressure broadening, modifying heat transport, amplifying the water vapor greenhouse feedback, and modulating atmospheric cooling via Rayleigh scattering. In addition, by exploring a broad parameter space of pressure and instellation within the Sparse Atmospheric Model Sampling Analysis (`SAMOSA`) Intercomparison, Haqq-Misra et al. (2022) demonstrated that `ExoPLASIM` produces climate states generally consistent with more complex GCMs, such as `ExoCAM`, while maintaining high computational agility. This validated tool was also employed by Macdonald et al. (2025) to evaluate the specific thresholds of atmospheric mass and transport discussed in Section 1.2.

### *1.3. Habitability indices*

Climate modelling enables different habitability criteria to be applied to the recovered 3D properties of the modelled planet, such as the temperature or precipitation distribution, yielding indices defined as the fractional surface area that meets the given criteria. To define the classical habitable zone, most works adopt the liquid water habitability index, hereafter $h_{lw}$, defined as the fraction of the surface with $T$ between the freezing and boiling points of water (e.g., Spiegel et al., 2008). A related index, $h_{50}$, introduced by Silva et al. (2017a), adopts a narrower temperature range, $0\ °C < T < 50\ °C$. This upper limit is suitable for the persistence of active metabolism, since it lies just below the temperature at which most proteins and nucleic acids denature (Nelson and Cox, 2004) and also lies near the threshold at which a runaway greenhouse may be activated (Kasting et al., 1993). In practice, as discussed in Silva et al. (2017a), the active metabolism temperature requirement was intended to target climates in which biology could persist long enough to affect atmospheric composition and produce biosignatures. Although this range was derived largely by constraints on ectothermic oxygen producers and consumers, it was argued to be broadly compatible with cyanobacteria as well.

Recent works have combined various metrics that leverage the extensive surface and atmospheric data provided by 3D models. These studies couple temperature-based metrics with measures of surface water availability, such as precipitation, evaporation, or evapotranspiration (Stevenson, 2019; Del Genio et al., 2019; Adams et al., 2025; Woodward, Rushby and Mayne, 2025). Such coupled metrics are particularly important for planets with continents, where land fraction and orography can strongly affect climate and habitability (Farnsworth et al., 2023; Laguë et al., 2023; Macdonald et al., 2022; 2025). These developments represent a step toward a more comprehensive view of habitability (e.g., Méndez et al., 2021).

### *1.4. This study*

Here we systematically assess the potential habitability of 22 planets listed in the Habitable Worlds Catalog (HWC[1]) using uniform assumptions across a range of planetary and atmospheric parameters. To this end, we utilize an adapted implementation of `PLASIM-LSG`—an intermediate complexity model coupling the `Planet Simulator` with the `Large-Scale Geostrophic Ocean Circulation Model` (Maier-Reimer et al., 1993)[2]—tailored to the specific conditions of tidally locked rocky planets orbiting M-type stars. This allowed us to simulate their climates, generate surface temperature maps, and evaluate their capacity to sustain liquid water. Modeling these systems as aquaplanets with surface pressures of $P_s = 0.5$, 1, and 5 $P_\oplus$, and a fixed $CO_2$ mixing ratio of 360 ppm, we simulate their climates, generate surface temperature maps, and evaluate their capacity to sustain liquid water. We quantify planetary habitability using the two temperature-based indices described above (the liquid-water index, $h_{lw}$, and the biologically oriented index, $h_{50}$) while also evaluating the precipitation- and evaporation-based index proposed by Woodward, Rushby, and Mayne (2025).

This climate model is particularly well suited for comparative studies over large exoplanet ensembles and extensive parameter spaces. As these planetary systems are expected to be observable with the next generation of space missions—such as the Habitable Worlds Observatory (HWO; National Academies of Sciences, Engineering, and Medicine, 2021) and the Large Interferometer For Exoplanets (LIFE; Quanz et al., 2022), and potentially even earlier with the James Webb Space Telescope through its 500-h Director's Discretionary Time "Rocky Worlds" program[3]—systematic investigations of the parameter space are essential to guide future observations by identifying the most promising candidates for hosting life.

---

[1] https://phl.upr.edu/hwc

[2] https://doi.org/10.5281/zenodo.4041461, with the modifications discussed in this paper currently available in the branch `devel_planet`.

[3] https://www.stsci.edu/contents/news/jwst/2024/stsci-initiates-a-concerted-search-for-atmospheres-around-m-dwarf-exoplanets

The remainder of this paper is organized as follows. In Section 2, we present the rationale for our selection of exoplanets. In Section 3, we describe our model, simulation protocol, and analysis methods. In Section 4, we present the main results, which are discussed in detail in Section 5. Concluding remarks and a summary of our findings are provided in Section 6.

## 2. Habitable Worlds Catalog

The HWC, previously known as the Habitable Exoplanets Catalog and currently maintained by the Planetary Habitability Laboratory at the Universidad de Puerto Rico at Arecibo, lists known Earth analogs with the highest Earth similarity index (ESI; Schulze-Makuch et al., 2011). ESI is a measure of how similar a planet is to Earth, quantified with the observable stellar instellation $S$ and the planetary radius $R$. It is defined as

$$\mathrm{ESI}\ (S, R) = 1 - \sqrt{\frac{1}{2}\left[\left(\frac{S - S_{\oplus}}{S + S_{\oplus}}\right)^2 + \left(\frac{R - R_{\oplus}}{R + R_{\oplus}}\right)^2\right]} \quad (1)$$

Although the full HWC catalog contains parameters for more than 5000 exoplanet candidates, in turn mainly compiled from the Planetary Systems Composite Table of the NASA Exoplanet Archive, we focused on Tables 1 and 2 therein. At the time of starting our analysis in 2024, Table 1 listed a conservative sample of 29 potentially habitable exoplanets with $0.5 < R/R_{\oplus} \leq 1.6$ and $0.1 < M/M_{\oplus} \leq 3.0$, while Table 2 listed an optimistic sample of 41 potentially habitable exoplanets with $1.6 < R/R_{\oplus} \leq 2.5$ and $3.0 < M/M_{\oplus} \leq 10.0$ (where $M$ is either the minimum mass only from radial velocity or the actual mass from both radial velocity and transits). Where direct measurements were unavailable, planetary masses or radii in the HWC were estimated from one another following Chen and Kipping (2017).

On the one hand, exoplanets in the optimistic HWC sample are less likely to have a rocky composition or to maintain surface liquid water than in the conservative sample, and some of them could turn out to be ocean worlds, hyceans, or mini-Neptunes (see Luque and Pallé, 2022, and references therein). On the other hand, all known Earth-analog candidates in the HWC conservative sample orbit around small (mostly M-type) stars. Therefore, such planets are very likely to be tidally locked (Tarter et al., 2007; Peale, 1977; Dole, 1964).

We went slightly beyond the conservative HWC sample and discarded seven exoplanets: those orbiting late K dwarfs at orbital distances unlikely to ensure tidal locking; planets with poorly constrained masses; and frozen planets whose parameter combinations prevent analysis with `PLASIM`[4]. These three additional filters ensured that our sample set consisted exclusively of telluric planets that probably are tidally locked to their host stars and yield meaningful results. As a result, we retained 22 Earth-like planets around 16 M dwarfs, as described in Section 3.

For each exoplanet, Table 1 lists the following: HWC ESI (in descending order); reported mass $M$ and radius $R$; derived surface gravity $g$; rotation period, $P$, which for tidally locked planets is also the orbital period; semi-major axis $a$; instellation, $S$, computed from $a$ and the stellar bolometric luminosity, $L_{\mathrm{bol}}$, via $S = L_{\mathrm{bol}} / 4\pi a^2$; emission temperature (effective radiating temperature for a blackbody) for a terrestrial Bond albedo, $T_{\mathrm{e}\ (A=0.3)}$; the global annual average surface temperature as calculated by `PLASIM`, $T_{\mathrm{s}\ \mathtt{PLASIM}}$. Values of $L_{\mathrm{bol}}$ are not reported, as they can be obtained from $S$ using the above-mentioned equation. Such $L_{\mathrm{bol}}$ were computed homogeneously by integrating the spectral energy distribution from the blue optical to the mid infrared, following the work of Cifuentes et al. (2020) though incorporating the latest stellar parallaxes and photometry from the Gaia Third Data Release. Updated values and additional details were sourced from Cifuentes (2023), the Transiting M dwarf Planets Catalog of Trifonov et al. (2021), and dedicated exoplanet discovery studies, such as those by Dreizler et al. (2024).

[4] The seven discarded planets are, respectively, Kepler-442 b, Kepler-62 f, LP 890-9 c, Kepler-186 f, GJ 667 C f, GJ 667 C e, and GJ 1002 c.

Table 1. Planetary data used in the simulations, ranked by descending ESI. [a]

| EXOPLANET | ESI | $M$ ($M_\oplus$) | $R$ ($R_\oplus$) | $g$ (m s$^{-2}$) | $P$ (d) | $a^{(b)}$ (au) | $S$ (W m$^{-2}$) | $T_{e\,(A=0.3)}$ (K) | $T_{s\,\mathtt{PLASIM}}$ (K) | Regime |
|---|---|---|---|---|---|---|---|---|---|---|
| Teegarden's Star b | 0.97 | 1.16 | 1.05 | 10.32 | 4.91 | 0.0259 | 1465 | 259 | 302 | 3 |
| TOI-700 d | 0.94 | 1.25 | 1.07 | 10.71 | 37.4 | 0.1633 | 1170 | 245 | 255 | 2 |
| Kepler-1649 c | 0.93 | 1.20 | 1.06 | 10.47 | 19.5 | 0.0649 | 1667 | 268 | 285 | 3 |
| TOI-700 e * | 0.91 | 0.82 | 0.95 | 8.91 | 27.8 | – | 1742 | 271 | 306 | 3 |
| TRAPPIST-1 d | 0.91 | 0.39 | 0.79 | 6.13 | 4.05 | 0.0223 | 1359 | 254 | 313 | 3 |
| K2-72 e | 0.87 | 2.21 | 1.29 | 13.02 | 24.2 | 0.1060 | 1769 | 272 | 284 | 3 |
| Proxima Cen b | 0.86 | 1.07 | 1.03 | 9.90 | 11.2 | 0.0486 | 924 | 231 | 236 | 2 |
| GJ 1002 b | 0.86 | 1.08 | 1.03 | 9.98 | 10.3 | 0.0457 | 916 | 231 | 234 | 2 |
| GJ 1061 d | 0.86 | 1.64 | 1.16 | 11.95 | 13.0 | 0.0540 | 794 | 222 | 220 | 2 |
| GJ 1061 c | 0.86 | 1.74 | 1.18 | 12.26 | 6.69 | 0.0350 | 1889 | 276 | 335 | 3 |
| Ross 128 b | 0.86 | 1.40 | 1.11 | 11.14 | 9.87 | 0.0496 | 2130 | 285 | 346 | 3 |
| GJ 273 b | 0.85 | 2.89 | 1.51 | 12.43 | 18.6 | 0.0911 | 1607 | 265 | 273 | 3 |
| Kepler-296 e | 0.85 | 2.96 | 1.53 | 12.40 | 34.1 | 0.1690 | 1287 | 251 | 255 | 2 |
| Wolf 1069 b * | 0.85 | 1.26 | 1.08 | 10.59 | 15.6 | – | 885 | 229 | 226 | 2 |
| TRAPPIST-1 e | 0.85 | 0.69 | 0.92 | 8.02 | 6.10 | 0.0293 | 788 | 222 | 229 | 2 |
| Kepler-1652 b * | 0.83 | 3.19 | 1.60 | 12.22 | 38.1 | – | 1143 | 244 | 241 | 2 |
| K2-3d * | 0.81 | 2.20 | 1.46 | 10.12 | 44.6 | – | 1973 | 279 | 313 | 3 |
| TOI-715 b * | 0.81 | 3.02 | 1.55 | 12.33 | 19.3 | – | 966 | 234 | 230 | 2 |
| TRAPPIST-1 f | 0.68 | 1.04 | 1.05 | 9.33 | 9.21 | 0.0385 | 455 | 194 | 204 | 1 |
| Teegarden's Star c | 0.66 | 1.05 | 1.02 | 9.90 | 11.4 | 0.0455 | 475 | 196 | 200 | 1 |
| Kepler-1229 b | 0.62 | 2.54 | 1.40 | 12.71 | 86.8 | 0.3006 | 574 | 205 | 199 | 1 |
| TRAPPIST-1 g | 0.58 | 1.32 | 1.13 | 10.16 | 12.35 | 0.0468 | 307 | 175 | 187 | 1 |

**Notes.** [a] ESI: Earth similarity index; $M$: planetary mass; $R$: planetary radius; $g$: surface gravity; $P$: revolution period; $a$: semimajor axis; $S$: stellar insolation; $T_{e\,(A=0.3)}$: emission temperature assuming a terrestrial albedo, $A_{Bond}$ = 0.30; $T_{s\,\mathtt{PLASIM}}$: surface temperature calculated by `PLASIM`. Where direct measurements were unavailable, planetary masses or radii in the HWC were estimated from one another following Chen and Kipping (2017). "Regime" refers to the classification of exoplanets as Snowball (1), Eyeball (2), or Hot (3), as discussed in Sect. 4.3. [b] Stellar insolations have been calculated as a function of semi-major axes and bolometric luminosities, with the only exception of five cases where the literature was not exhaustive or discordant. In the latter cases, marked by an asterisk, stellar insolations are those of the HWC. The semimajor axes were taken from Agol et al., (2021); Anglada-Escudé et al. (2013); Astudillo-Defru et al. (2017); Barclay et al. (2015); Bonfils et al. (2018); Borucki et al. (2013); Delrez et al. (2022); Dreizler et al. (2020); Dreizler et al. (2024); Dressing et al. (2017); Faria et al. (2022); Gilbert et al. (2023); Suárez Mascareño et al. (2023); Torres et al. (2015); Torres et al. (2017).

Figure 1 shows planetary mass as a function of the orbital (rotational) period, illustrating the parameters listed in Table 1. The 22 exoplanets in our sample set possess radii ranging from 0.79 to 1.55 $R_\oplus$, with average surface temperatures, simulated by `PLASIM`, falling between approximately 190 and 330 K. The conservative sample set of habitable planets of the HWC may expand in the near future with the addition of newly discovered Earth-sized planets in the habitable zone (e.g., GJ 12 b; Kuzuhara et al., 2024; Turner et al., 2026). Ultimately, our choice of the ESI-based catalog ensures a homogeneous sample of observed planets, adhering to a well-defined metric (the ESI) and eliminating the potential for arbitrary target selection.

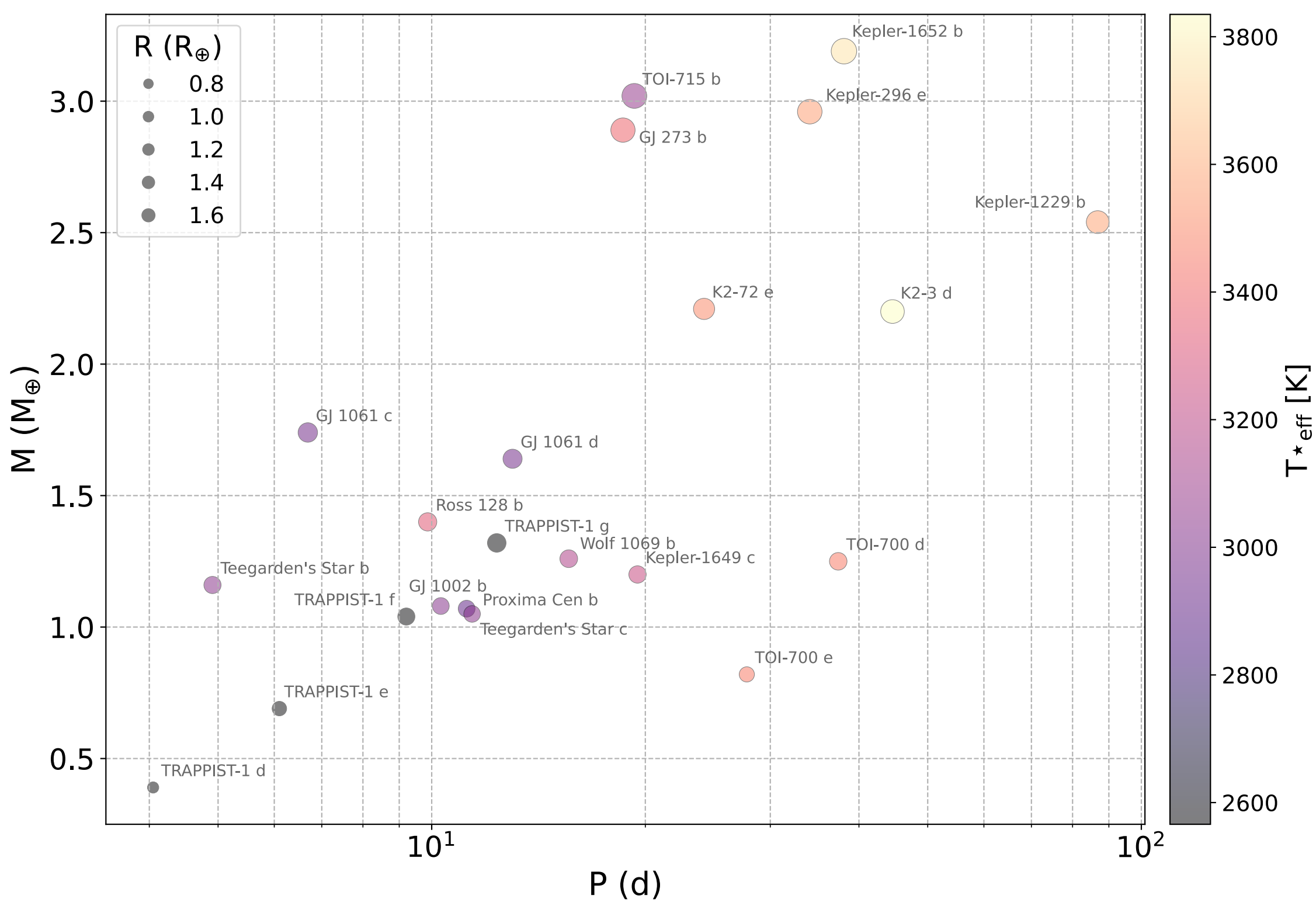

**Figure 1**. Planetary mass as a function of orbital (rotation) period. Symbol size is proportional to the planetary radius, while the color scale indicates the host star's effective temperature ($T^{*}_{\mathrm{eff}}$). Values are sourced from Table 1. Effective temperatures were obtained from Stassun et al. (2019) and Marfil et al. (2021).

## 3. Methods

### *3.1. The* `PLASIM` *climate model*

To systematically simulate, classify, and compare the climates of potential habitable exoplanets across diverse parametric configurations, we adopted the intermediate complexity global climate model `PLASIM` (`PLAnet SIMulator`; Fraedrich et al., 2005). This model includes an atmospheric dynamical core, namely the numerical component that simulates the three-dimensional, large-scale circulation of the atmosphere and the evolution of the winds, temperature, pressure, and moisture transport. It also incorporates simplified parameterizations of surface-atmosphere exchanges, radiation, moist processes, clouds, convection, and land-surface processes, thereby representing the main components of the hydrological cycle.

The model code used in this work (`PLASIM-LSG`; Hertwig et al., 2015; Angeloni et al., 2020; Mehling et al., 2023) included a configuration with a fully three-dimensional dynamic ocean component (`LSG`; Maier-Reimer et al., 1993), though we reduced model complexity by simulating sea-surface temperatures with a slab-ocean component of fixed shallow depth, and with horizontal oceanic heat transport parameterized by the addition of a horizontal diffusion term (Lunkeit et al., 2011). This approach approximated a well-mixed oceanic surface layer and captured its main thermodynamic properties. Sea ice was simulated using a simple thermodynamic model, excluding transport, based on the zero-layer model by Semtner (1976). The inclusion of a dynamical core required spectral integration.

An important advantage of using an intermediate complexity model like PLASIM for this type of study is that, while fully three-dimensional, it includes fewer details compared to state-of-the-art full global climate models typically used to study Earth's climate evolution (e.g., detailed geography and topography, including localized physical processes) and employs fewer and simpler parameterizations with a reduced number of parameters. As such, it is suitable for application to exoplanets, for which most of the detailed information required for a full GCM is currently unknown and will likely remain so in the near future. A second advantage is that PLASIM, in the configuration adopted for this study, is computationally efficient; a simulation equivalent to a single orbital period typically runs on a single core in under one minute. This performance enables rendering of large ensembles, which facilitates a comprehensive exploration of parameter spaces.

Although originally developed for Earth's climate, different configurations of PLASIM have also been adapted and used for planetary studies, including investigations of solar system planets (Grieger et al., 2004; Segschneider et al., 2005), exploration of Snowball-Earth climate regimes (Fraedrich, 2012), and the study of exoplanetary atmospheres and habitability. In particular, variations of PLASIM have been used to study the impact of obliquity and eccentricity (Jernigan et al., 2023), the atmospheric dynamics of a tidally locked planet (Galuzzo et al., 2021), and the impact of atmospheric pressure for planets in the habitable zone (Paradise et al., 2021; Macdonald et al., 2025).

While the simplifications inherent in intermediate complexity climate models make them suitable for studying a large sample set of exoplanets with a range of planetary climates, and they are expected to capture the main climate and habitability trends even for conditions far from those on Earth, a deeper understanding of subtle processes and small-scale feedbacks may still require investigation with full 3D GCMs. Regarding PLASIM-LSG specifically, since our modified version employed in this work is aligned with the ExoPLASIM code (as described in the following Section 3.2), it shares the same limitations, which have been comprehensively discussed by Paradise et al. (2022). Among these, we recall in particular the use of a simplified radiative transfer scheme based on only three spectral bands and the occurrence of Gibbs phenomenon artifacts that can affect the model output. To cope with the latter, we did not apply spectral filtering in the Fourier computations as was done, for example, by Paradise et al. (2022), but instead performed a resolution test to assess the influence of this numerical artifact. However, in Section 5, we compare our findings with those of other research groups and show that our results are within the variance that can be found in the literature. We also note that PLASIM does not include a treatment of photochemistry. As a result, it cannot model processes such as ozone formation or assess the long-term potential for water loss via hydrogen escape, both of which may be relevant for evaluating planetary habitability.

### *3.2. Code modifications*

We modified the PLASIM-LSG code to enable simulations of planetary bodies with varying radius, gravity, surface pressure, and orbital parameters, including the ability to model 1:1 tidally locked configuration. Furthermore, to allow simulations under non-solar stellar spectra, we adapted and integrated three key modifications into the radiation scheme. These spectral adjustments—specifically governing the energy partition between shortwave bands, Rayleigh scattering, and surface albedos—were originally introduced by Paradise et al. (2022) for ExoPlasim, a different version of the PLASIM code that lacks 3D ocean coupling. We refer to the reader to that paper for further technical details and an extensive comparison with other models with regard to the impact of various radiative parameterizations.

3.2.1. Energy partition between shortwave bands. The partitioning of the incident stellar flux between the two short-wave bands in PLASIM's radiation scheme (visible and near-infrared radiation at wavelengths shorter and longer than 0.75 μm, respectively) needed to be adapted as a function of the emission temperature of the host star. To this end, we adopted the routine included in ExoPLASIM (Paradise et al., 2022). When applied to a solar spectrum, this calculation provided a partition identical to the PLASIM default.

3.2.2. Rayleigh scattering. The parameterization of Rayleigh scattering for diffuse and direct beams at the surface as a function of the cosine of the solar zenith angle in PLASIM was originally developed for Earth's atmosphere and a solar spectrum. We adopted the parameterization introduced by Paradise et al. (2021) and further developed by Paradise et al. (2022), which accounted for the dependence of Rayleigh scattering on surface pressure, gravity, and the blackbody temperature of the host star.

3.2.3. Surface albedos. Due to the different energy partitioning between the two short-wave bands, the albedo for water, land, snow, sea ice, and glacial ice had to be recomputed for each host star spectrum. Following the work of Paradise et al. (2022), we computed separate albedos for both bands. For the simulations presented in this paper, we adopted the simplification of combining the albedos from both bands into a single bolometric albedo for each surface type.

### *3.3. Simulation protocol*

As is customary in this type of exploratory study, where only planetary mass and radius are known, we investigated the climate of the selected planets by assuming an Earth-like atmospheric composition and varying the surface pressure. Specifically, we assumed a fixed composition with a $CO_2$ volume mixing ratio of 360 ppm (corresponding to Earth's measured concentration in 1995), utilizing planetary configurations that include an aquaplanet setup, a 50-m deep slab ocean, and zero obliquity and eccentricity. Each planet in our sample set was assumed to be tidally locked in a 1:1 spin-orbit resonance. While this configuration is highly probable, there is currently no observational evidence to validate this assumption for the cases considered. Nevertheless, it enabled us to maintain a uniform setup across all simulations. The choice of an aquaplanet geography was not intended to represent planets possessing a literal global 50-m deep ocean, but rather it served as a model simplification to bypass uncertainties regarding landmass distribution and topography. Furthermore, this setup ensured that the climate system was not constrained by limited water availability. The presence of landmasses on tidally locked planets is known to significantly influence the climate (e.g., Macdonald et al., 2022; 2025). Because the ocean is represented as a 50-m slab (i.e., without resolved ocean dynamics), we interpreted habitability using surface climate diagnostics (e.g., near-surface temperature and sea-ice cover) rather than the temperature of a mixed-layer ocean. For small land fractions, the large-scale circulation and the first-order surface climate are expected to remain close to the aquaplanet limit.

We performed three sets of simulations: (*i*) with Earth surface pressure $P = P_\oplus$, (*ii*) with $P = 0.5P_\oplus$, and (*iii*) with $P = 5P_\oplus$. Since the $CO_2$ mixing ratio was held constant, this also resulted in a variation of the total greenhouse gas mass, with lower pressures implying less $CO_2$ and higher pressures implying more. The chosen range of surface pressures spans a full order of magnitude, which allowed us to investigate the climate response across a broad parameter space. This was especially relevant for our sample set of planets, which spans a wide range of planetary radii, gravities, and—most notably—rotation periods. We note that, under this simplifying assumption, often adopted in the absence of stronger constraints, the atmospheric mass scaled proportionally with planetary mass (e.g., Kopparapu et al., 2014; Silva et al., 2017b), and the expected surface pressures for our sample set ranged from approximately 0.4 to 1.7 bar. Our chosen pressure range thus encompassed and extended beyond this interval and ensured that we captured both realistic and boundary-case scenarios.

In addition, in this study we aimed to maintain the highest possible degree of homogeneity across our sample by varying only one parameter at a time. Specifically, for each planet we included only observed quantities—mass and radius—while holding all other parameters constant, including atmospheric composition. This also motivated our choice to compare climates at fixed surface pressures. Varying surface pressure arbitrarily would have altered atmospheric heat transport in ways unrelated to planetary observables. By maintaining a constant pressure across the entire sample set, we ensured that differences in climate behavior arose solely from known planetary parameters, which allowed us to isolate the impact of each of them.

Planetary radius, surface gravity, semi-major axis, orbital period, stellar irradiance, and stellar blackbody temperature were set according to the values reported in Table 1 and Figure 1. Other parameters, such as the horizontal heat diffusion coefficient in the slab ocean, were set to default values used for Earth. All these runs were performed at low resolution (according to `PLASIM` terminology, resolution T21, that is, 64 longitudinal and 32 latitudinal points). The robustness of the results at this numerical resolution was assessed by repeating some simulations at double resolution (see Section 4.1). Crucially, in our `PLASIM` simulations we employed 19 vertical levels. This decision was motivated by the intense atmospheric convection characteristic of synchronously rotating planets. Since `PLASIM` automatically determines the tropopause level, defined as the layer where parametrized convection drops to zero, the standard 10-level resolution may fail to resolve these dynamics. A coarser vertical grid might not resolve the tropopause and lead to inaccurate simulations, as the model would deactivate the convection parametrization only above this level. In our setup, these vertical levels were defined in terms of the σ coordinate ($\sigma = P/P_s$), with 19 levels ranging from 0.0007 to 1.0000 following an approximately logarithmic spacing. Depending on the planetary gravity, the highest level (corresponding to the lowest σ value) reached an equivalent altitude between approximately 40 and 90 km.

The simulations were integrated for a duration equivalent to approximately 100 terrestrial years, with the exact number of `PLASIM` years (each defined as 365 model days[5]) adjusted according to each planet's orbital period. In all cases, climate data were averaged over the final 365 planetary orbits. For example, TRAPPIST-1 d (period = 4.05 days) was integrated for 25 `PLASIM` years, equivalent to ∼101 terrestrial years. TOI-700 d and K2-3 d were simulated for 3 `PLASIM` years (∼112 and ∼133 terrestrial years, respectively), while Kepler-1229 b—given its longer period of 86.8 days—was run for 2 `PLASIM` years (173 terrestrial years). We verified that the surface temperature had stabilized during this period, confirming that the planets had reached thermal equilibrium between incoming stellar radiation and outgoing emission. Simulations were initialized from an atmosphere at rest with a default initial temperature of 320 K, following a standard "hot start" configuration.

## 4. Results

### 4.1. Surface temperature

A peculiarity of tidally locked planets is that their revolution period is fixed by the separation from the host star, and inclination and eccentricity are driven toward a value of zero by the tidal force. As a consequence, the size of the parameter space is reduced with respect to non-tidally locked planets, and stellar flux becomes the most important factor that impacts habitability. With zero eccentricity and inclination, hence no seasonal variability, the global average surface temperature becomes the most appropriate proxy for assessing habitability. However, the physical rotation speed of a planet does have an impact on its climate via the Coriolis force.

[5] We retained the default `PLASIM` calendar configuration. Under the assumption of synchronous rotation for our targets, one `PLASIM` day is equivalent to one full orbital period; consequently, a standard `PLASIM` year of 365 days represents 365 individual orbits.

To characterize the dynamical regimes of our planetary sample set, we adopted the classification framework proposed by Haqq-Misra et al. (2020). We defined the equatorial Rossby radius of deformation, $\lambda_r$, as

$$\lambda_r^2 = \frac{\sqrt{gH}}{2\beta} R_p \tag{2}$$

where $g$ is the planetary gravitational acceleration, $R_p$ is the planetary radius, and $H = (T_s \cdot R) / (g \cdot M_{air})$ is the atmospheric scale height (with $T_s$ being the surface temperature, $R$ = 8.314 J mol$^{-1}$ K$^{-1}$ the universal gas constant, and $M_{air}$ = 28.9644 × 10$^{-3}$ kg mol$^{-1}$ the molar mass of air). To account for the varying sizes and spherical geometries of our diverse planetary sample set, we explicitly used the spherical coordinate definition for the equatorial Coriolis parameter gradient, $\beta = 2\Omega/R_p$ (where $\Omega$ represents the planet's angular velocity), thereby maintaining strict dimensional consistency across the dataset (Matsuno, 1966). Additionally, the Rhines length was defined as

$$L_r = \pi\sqrt{\frac{U}{\beta}} \tag{3}$$

where $U$ represents the characteristic root-mean-square horizontal velocity. Under this scheme, a planet is classified as a *slow rotator* if $L_r/R_p > 1$, a *fast rotator* if $L_r/R_p < 1$, and a *Rhines rotator* if $L_r/R_p < 1$ but $\lambda_r/R_p > 1$. Based on these criteria at the 1-bar pressure level, our sample set was comprised of four fast rotators (Teegarden's Star b, TRAPPIST-1 d, GJ 1061 c, and TRAPPIST-1 e) and four Rhines rotators (Ross 128 b, TRAPPIST-1 f, Teegarden's Star c, and TRAPPIST-1 g), while the remaining planets are classified as slow rotators.

Figure 2 illustrates the surface temperature (the atmospheric temperature at a height of 2 m above the surface in each cell) of the 22 exoplanets. The maps are centered on the substellar point. Three climate regimes can be identified: Hot climates with a moderate temperature gradient between the two hemispheres, "Eyeball" climate states with a temperate substellar circular region and ice covering in the surrounding regions, and fully ice-covered Snowball climate states. Those with a significant "Eye" have a substellar region in terms of diameter at $T > 273.15$ K, while other regions have a more uniform temperature, with low or no habitability. For example, according to our definition of habitability (Spiegel et al., 2008), TOI-700 d is habitable only in the region surrounding the substellar point, while the remaining portion of its surface is too cold, with its temperature dropping below the freezing point of water. This agrees with the results found in other studies on tidally locked planets (e.g., Lobo and Shields, 2024). However, three planets (Kepler-1649 c, K2-72 e, GJ 273 b) exhibited an intermediate behavior, with part of their night side free of ice. We refer to these worlds as "Transitional." A similar configuration was identified by Macdonald et al. (2025), who also referred to it as a "transitional regime." We also note that, based on our computations, Ross 128—the planet with the highest insolation in our sample set (2130 W m$^{-2}$, 1.58 times the solar constant)—exhibited a surface temperature in the range of approximately 67 °C to 80 °C. Consequently, it should not be considered a truly habitable world. Both exceeded the 50 °C temperature limit for complex life and are likely prone to the onset of a runaway greenhouse effect.

In the present work, we focused on the surface temperature and the presence of sea ice. However, we also found that the substellar region of Eyeball climates is characterized by the presence of clouds, convection, and precipitation, with a hydrological cycle mainly concentrated in that region: precipitations are concentrated in the warm region and are practically absent on the dark side, as already found by Lobo and Shields (2024). This holds true across all dynamical regimes, even though the distribution of humidity and clouds can be influenced by fast rotation and slightly shifted relative to the substellar point, as discussed in Section 4.3. This effect can also be observed in Figure 4, which includes TRAPPIST-1 d, the fastest rotator in our sample.

For Hot planets, the situation was found to be different: precipitation was often present on the dark side, especially at high latitudes, and cloud cover was much more uniform, with several planets having smaller cloud cover in the substellar region. Their albedo is consequently lower as well, which helps warm up the entire planet. In general, temperatures in the dark hemisphere, although subzero, remained above the condensation points of atmospheric $O_2$ and $N_2$, which prevented the risk of atmospheric collapse. A more detailed discussion of this point is provided in Section 4.3.

To assess the robustness of our results, we also tested the sensitivity to model resolution. We repeated the simulations for TOI-700 d and TRAPPIST-1 d at double numerical resolution, using T42 instead of T21. The results showed excellent agreement: in particular, the global averages differed by no more than 5%. As for the Gibbs phenomenon, that is, the appearance of spurious oscillations ("waves") in physical variables due to numerical artifacts, Paradise et al. (2022) mitigated this numerical artifact by applying a filter to the spectral computations. We note that a similar reduction can also be achieved by increasing the model resolution. This effect is particularly evident in the distribution of clouds on the night side of the planet, where the clouds themselves play a relatively minor role.

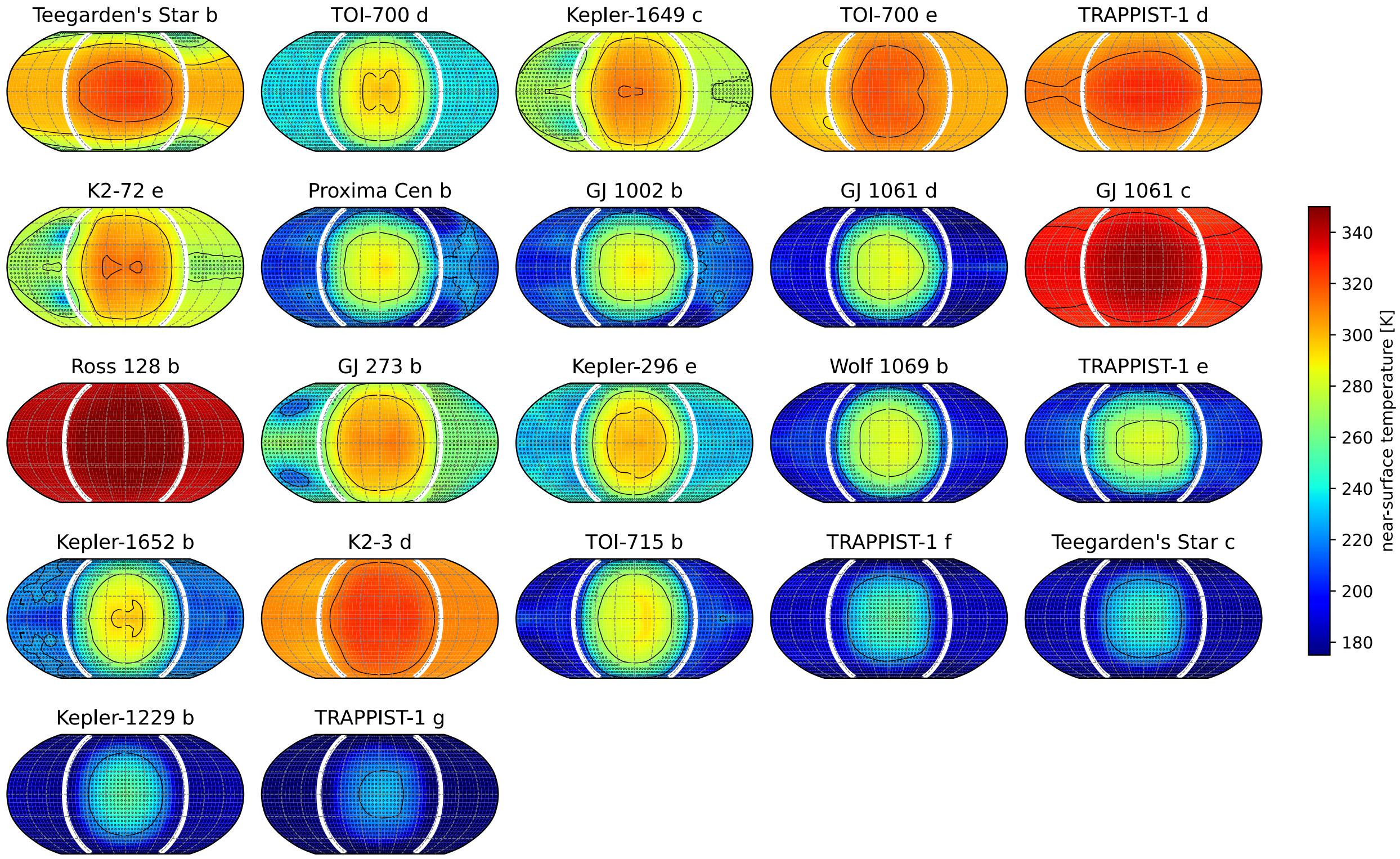

**Figure 2**. Surface temperature of the 22 exoplanets in an aquaplanet configuration. Obliquity and eccentricity are set to zero, and all are tidally locked with Earth-like atmospheric chemistry and pressure. Level lines correspond to the temperatures of 223.15, 273.15, and 323.15 K. White lines indicate the terminator, that is, the boundary between the illuminated and the dark sides of a planet. Black dots indicate the prevalence of sea ice (ice fraction larger than 0.5).

### 4.2. Effect of pressure

Since Figure 2 shows that some planets are uninhabitable, due to either a hot climate state or extensive ice cover, we further explored the dependence of habitability on surface pressure by testing values both below and above the default Earth-like pressure of 1 bar. Although the dependence of tidally locked climates on surface pressure was analyzed in detail by Paradise et al. (2021) with a fixed $CO_2$ partial pressure, here we investigated the effect of varying the total pressure while keeping the $CO_2$ mixing ratio constant at its default value of 360 ppm. Note that this meant changing the actual $CO_2$ mass in the atmosphere.

Figure 3 displays the effect of varying the surface pressure on five planets: TOI-700 d, TRAPPIST-1 d, K2-72 e, TRAPPIST-1 e, and K2-3 d. Specifically, TOI-700 d (slow rotator) and TRAPPIST-1 e (fast rotator) are classified as Eyeball planets. K2-3 d (slow rotator) and TRAPPIST-1 d (fast rotator) are classified as Hot planets. Finally, K2-72 e is a Transitional, slow-rotating planet. In the figure, they are presented in order of ESI, from the highest to the lowest.

In addition to the reference pressure value, we considered two alternative surface pressures, $P_s = 0.5\ P_\oplus$ and $P_s = 5\ P_\oplus$. The habitability of Eyeball planets was generally not strongly affected by changes in pressure and the associated increase in $CO_2$ mass—at least within the range considered. In contrast, the habitability of Hot planets in terms of $h_{50}$ benefited from low pressures and thus low $CO_2$ mass (at a fixed fraction). Given that we changed the surface pressure by keeping the atmospheric compositions fixed, lower-pressure planets were generally colder, while they were hotter at higher pressure. This is in part due to the fact that the absolute amount of $CO_2$ was smaller at lower pressures, which thus reduced the importance of the greenhouse effect. The opposite situation occurs at higher pressure. Our results also confirmed a well-known effect: higher pressures tend to favor heat transfer and reduce temperature gradients, while the opposite happens at lower pressure (cf. Paradise et al., 2021). This finding is largely independent of the rotation speed. Only the Hot fast rotator TRAPPIST-1 d showed a significant latitudinal distortion in the temperature distribution, due to the presence of strong zonal winds. Note that this feature is far less pronounced for TRAPPIST-1 e, which is also a fast rotator but is classified as an Eyeball planet.

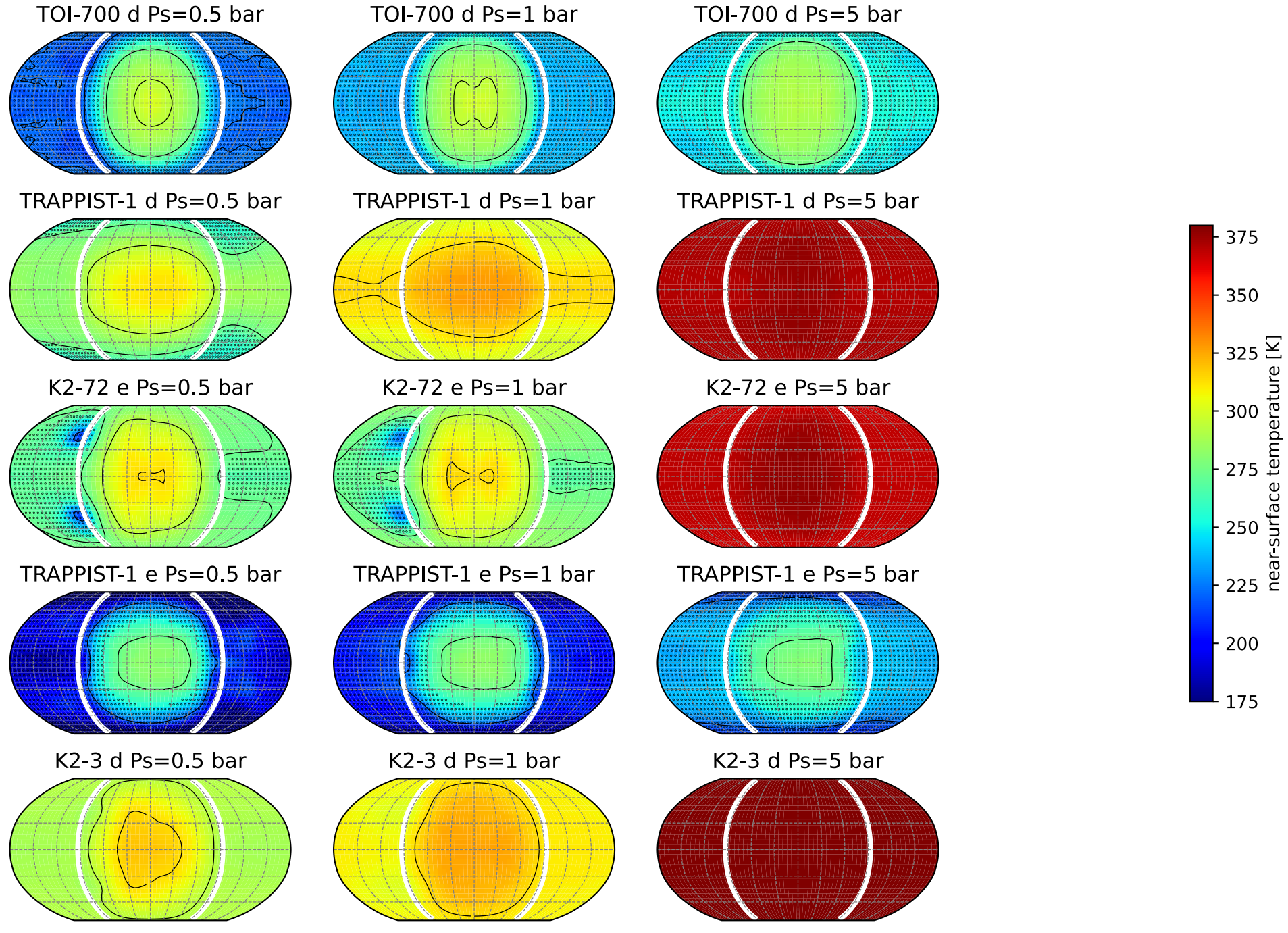

**Figure 3**. Same as Figure 2 but for three distinct surface pressures and five representative exoplanets, two exhibiting an Eyeball climate (in slow- and fast-rotating configurations respectively), two exhibiting a Hot climate (in slow- and fast-rotating configurations), and one in the Transitional state. From top to bottom: TOI-700 d, TRAPPIST-1 d, K2-72 e, TRAPPIST-1 e, and K2-3 d. Left maps: $P_s = 0.5\ P_\oplus$; central maps: $P_s = P_\oplus$; right maps: $P_s = 5\ P_\oplus$.

### *4.3. Vertical structure of the atmosphere*

We investigated the vertical atmospheric structure for the same representative five planets considered in Figure 3. Figure 4 displays equatorial cross-sections of temperature, specific humidity, cloud cover, and cloud liquid water content, derived from our simulations at a surface pressure of 1 bar. These plots span all longitudes from -180° to 180° and represent local values rather than zonal averages. The bottom row reproduces the near-surface temperature maps shown in Figure 2 and Figure 3 for the respective planets. Vertical levels are expressed in terms of the sigma coordinate, $\sigma = P/P_s$ (see Section 3.3).

The vertical temperature structure of the Eyeball planets is straightforward: the highest temperatures correspond to the position of the "Eye" (substellar region); low temperatures were found even at high σ values. Note the thermal inversion near the surface, outside the Eye. The Hot planets exhibited similar behavior but reached significantly higher temperature scales. The temperature contours of the fast rotator TRAPPIST-1 d are distorted in the direction of rotation, due to zonal winds. No inversion was present, and low temperatures (well below 0 °C) were reached only at low σ levels. The Transitional planet K2-72 e showed a slightly more complicated structure due to the fact that, even when we considered equatorial cross-sections, we found on the night side an alternation of frozen and ice-free regions. Specific humidity closely followed the pattern of vertical temperatures. Hot planets have a larger amount of water vapor in their atmospheres, and significant humidity was present also on their night side. Eyeball planets, instead, have almost no humidity outside the position of the Eye, and the humidity regime of a Transitional planet falls in between that of Hot and Eyeball planets. The strong transport of water vapor driven by zonal winds was clearly evident for TRAPPIST-1 d.

In all cases except for TRAPPIST-1 d, cloud cover was strongly concentrated at the position of the Eye. TRAPPIST-1 e, a fast Eyeball rotator, exhibited some cloud cover over the eastern night side, advected by weak zonal winds. The strong zonal winds of TRAPPIST-1 d distributed its clouds across all longitudes. However, comparing the cloud cover with the cloud water content, we note that, while the highest cloud fraction was found at high altitudes, the bulk of the water content resided at low altitudes. The simulations appear to distinguish between high-altitude and low-altitude clouds. In all cases except for TRAPPIST-1 d, the cloud water content, and thus precipitation (not shown), was concentrated at the position of the Eye. TRAPPIST-1 d, in agreement with the water vapor distribution, exhibited precipitation that is shifted westward.

All other planets in our sample exhibited behavior similar to those described here, depending on their temperature and rotation class. A notable outlier was Ross 128 b, which—due to its very high incident stellar flux—featured an atmosphere that was fully saturated with water vapor up to the highest simulated atmospheric level. GJ 1061 c, the second hottest planet in the sample set, also featured a large amount of humidity at high altitudes but did not appear to be completely saturated. Fast rotators all exhibited a characteristic deformation of their cloud and hydrological systems driven by intense zonal winds. Rhines rotators consistently emerged as climatic outliers; this specific group comprises Ross 128 b alongside three planets in a Snowball state. Compared to Eyeball configurations, all Hot planets featured a reduced cloud fraction while maintaining substantial specific humidity, particularly at low altitudes. Conversely, we found that Eyeball planets are characterized by a hydrological cycle tightly concentrated within the substellar region. It is worth noting that cloud cover was averaged over 365 orbits to ensure consistency with all other physical quantities presented in this study. Where the atmospheric system is stable, cloud cover can exceed 90%. In contrast, clouds on Hot planets displayed more variable, transient patterns, generally resulting in a more widely distributed but lower time-averaged coverage (with four notable exceptions discussed in Section 4.4).

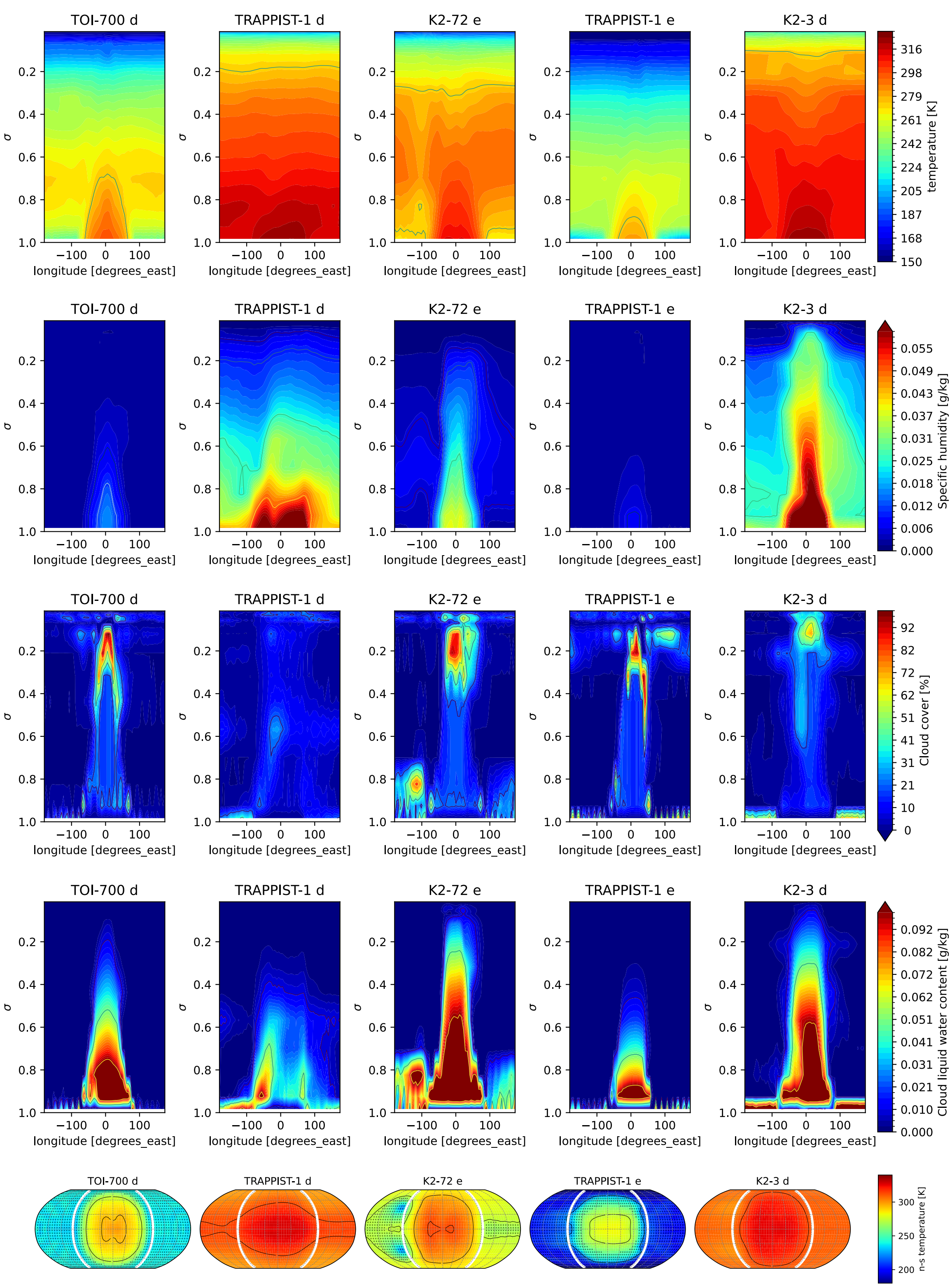

**Figure 4**. From top to bottom: Equatorial cross-sections of temperature, specific humidity, cloud cover, and cloud water content, derived from our simulations at a surface pressure of 1 bar for a representative sample of exoplanets. These plots span all longitudes from -180° to 180° and represent local values rather than zonal averages. The bottom row reproduces the near-surface temperature maps shown in Figure 2 and Figure 3 for the respective planets. Vertical levels are expressed in terms of the sigma coordinate, $\sigma = P/P_s$.

From a thermodynamic perspective, increasing the atmospheric pressure under isothermal conditions reduced the specific humidity; indeed, the saturation specific humidity is inversely proportional to the total pressure. Moreover, a higher atmospheric mass dampened both convection and horizontal thermal gradients. As a consequence, specific humidity and cloud water content decreased as pressure increased. This trend was observed in our simulations of Eyeball planets. Conversely, hotter planets reached extremely high surface temperatures at 5 bar owing to the increased $CO_2$ mass. In this regime, enhanced evaporation became the dominant effect, driving the atmosphere toward water vapor saturation—a behavior that points to a potential runaway greenhouse effect. Indeed, across all simulated cases, a direct consequence of increasing the pressure at a fixed chemical composition was a net increase in atmospheric temperature at all altitudes. As a result, the zero-point altitude also shifted upward.

### *4.4. Day-side–night-side energy balance*

Figure 5 displays the energy flux balance between the day and night hemispheres for the selected set of 22 exoplanets, sorted by increasing equilibrium temperature. The x-axis represents the averaged energy flux in watts per square meter. Planets are sorted in order of increasing instellation ($S$). For each planet, we report the averaged energy flux absorbed by the illuminated hemisphere, as well as the averaged thermal energy flux irradiated by it. We observed an approximately linear increase in the map-averaged net radiation of the illuminated hemisphere ($\langle r_{\text{net}} \rangle$) with increasing instellation, which acts as the primary driver of atmospheric dynamics. The relationship can be expressed as $\langle r_{\text{net illuminated}} \rangle \geq 0.2\, S - 62.8$ W m$^{-2}$. Similar relationships can be found for the other quantities plotted in the figure, with the exception of the total net radiation $\langle r_{\text{net total}} \rangle$.

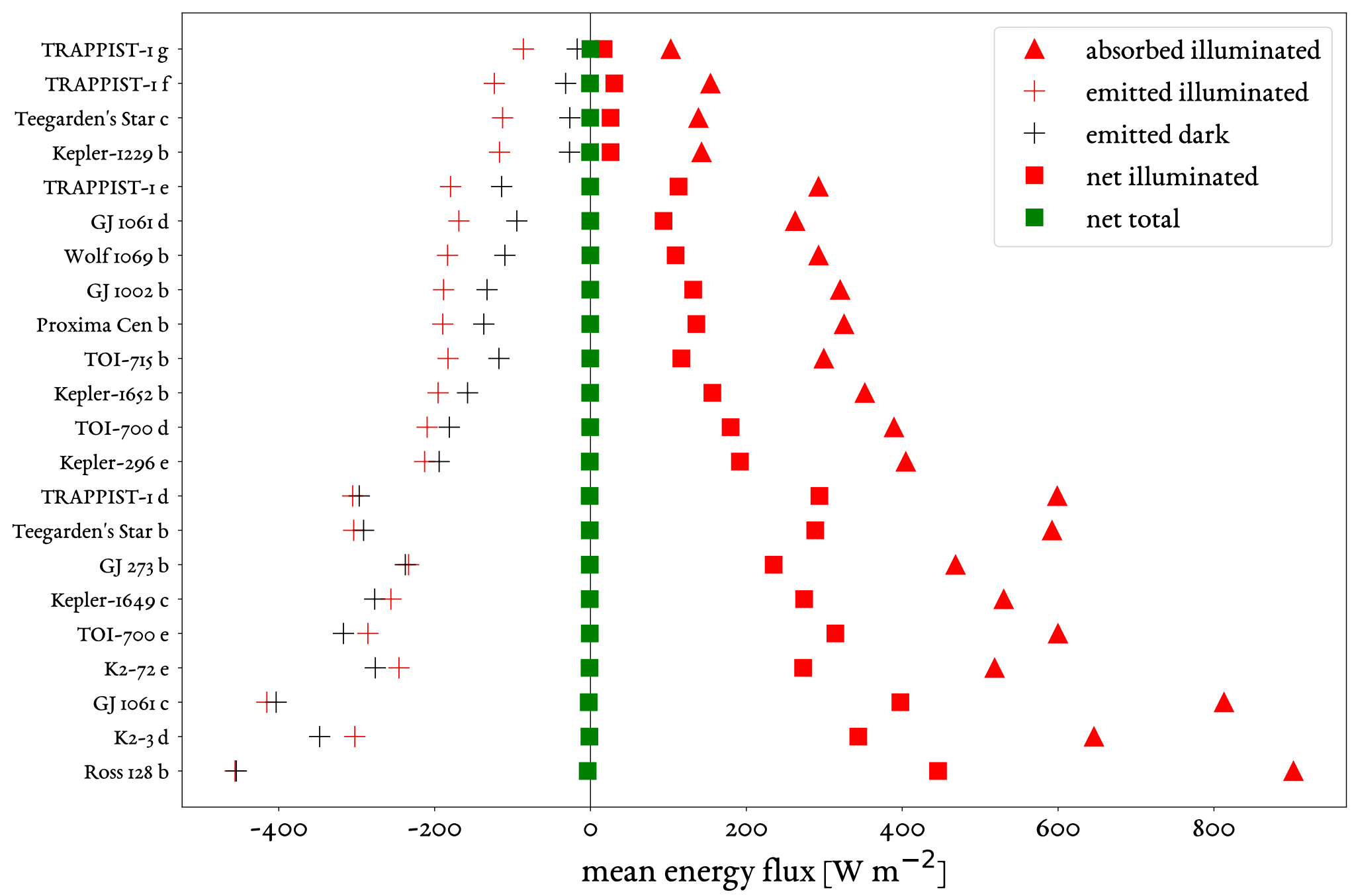

**Figure 5**. Balance of averaged radiative energy fluxes for the illuminated and dark hemispheres. Red

symbols represent quantities for the illuminated side: mean absorbed radiation (triangles), mean emitted radiation (crosses), and mean net radiation (squares; the sum of absorbed and emitted components). Dark-side emission is indicated by black crosses. Green squares denote the net radiation averaged over the entire planetary surface. Planets are ordered by increasing instellation.

Following a common convention in climate modelling, incoming (downward) energy fluxes are positive, while irradiated (outgoing) fluxes are negative. Since energy flux absorption on the dark hemisphere is zero, it is not displayed. In most cases, averaged thermal energy fluxes on the dayside exceeded those on the nightside. The averaged net energy flux for each hemisphere is defined as the sum of its averaged absorbed and irradiated energy. Given that absorption on the dark hemisphere is null, its averaged net energy flux coincided with its averaged thermal irradiated flux. The averaged net energy flux on the illuminated side was consistently positive, as a portion of the incoming radiation was absorbed. Consequently, the averaged net energy flux on the dark side must balance this excess, ensuring the total averaged net flux (the average of both hemispheres, green squares) equals zero.

Notably, for all non-Hot planets, the averaged radiative flux emitted from the illuminated side was higher than from the dark side. Conversely, Hot planets exhibited comparable emissions from both hemispheres, or even a slightly higher emission from the dark side. The energy flux absorbed by the illuminated side and subsequently transferred to the night side ranged between 17 and 32 W m$^{-2}$ (approximately 16.5% to 20.7% of the total absorbed energy flux) for planets in the Snowball regime. This efficiency increased significantly for Eyeball states, with heat transfers between 94 and 194 W m$^{-2}$ (36% to 48%), and reached its peak in Hot planets, where the transfer ranged from 274 to 454 W m$^{-2}$ (49.2% to 53.8%).

Four Hot planets—TRAPPIST-1 d, Teegarden's Star b, GJ 1061 c, and Ross 128 b—deviated significantly from the aforementioned linear relationship; they exhibited a notable excess in net radiation. Inspection of their cloud-cover and albedo maps reveals a markedly different spatial distribution compared with that of the Hot planets, which followed a linear trend. In the latter group, cloud cover on the illuminated hemisphere extended across nearly all latitudes below approximately 50°–60°, with a slight preference for positive longitudes. In contrast, the four anomalous planets displayed cloud cover concentrated within a narrow equatorial band. This specific cloud distribution resulted in a substantially lower planetary albedo, which caused a larger fraction of the incident stellar radiation to be absorbed and subsequently re-emitted. Consequently, these planets exhibited higher net radiation than expected from the general trend. This discrepancy arises because three out of these four planets are fast rotators, and the fourth (Ross 128 b) is a Rhines rotator. As shown by Haqq-Misra et al. (2018), strong zonal winds in these dynamic regimes limit the meridional transport of moisture, thereby suppressing cloud formation far from the equator. Conversely, all other Hot planets in our sample set qualify as slow rotators.

Regarding heat transport, we found that stellar flux is the primary driver of climate, while the rotational dynamical regime played a less significant role—at least within our sample set. Among the four fast rotators, we observed both Hot and Eyeball climates, whereas our Rhines rotators exhibited Hot and Snowball regimes. However, because the majority of the planets in our dataset are classified as slow rotators, we cannot draw a firm conclusion regarding the link between rotation and climate regimes.

### *4.5. Classification of exoplanets*

The three climate regimes introduced in Section 4.1 and illustrated in Figure 2 can be better explored by quantifying the prominence of the Eyeball climate configuration. To this end, we considered the mean surface temperature $T_s$, defined as

$$T_s = \frac{1}{4\pi} \int_{4\pi} T(\hat{\Theta})\, d^2\hat{\Theta} \tag{4}$$

where $\hat{\Theta}$ denotes the surface location of the planet expressed as the usual latitude and longitude coordinates. The difference in surface temperatures between the substellar ($\theta = 0°$) and antistellar ($\theta = 180°$) points is defined as

$$\Delta T_{sa} = T(\theta = 0°) - T(\theta = 180°) \tag{5}$$

which represents a measure of the temperature variation range between the two hemispheres (the subscript "sa" stands for "stellar-antistellar").

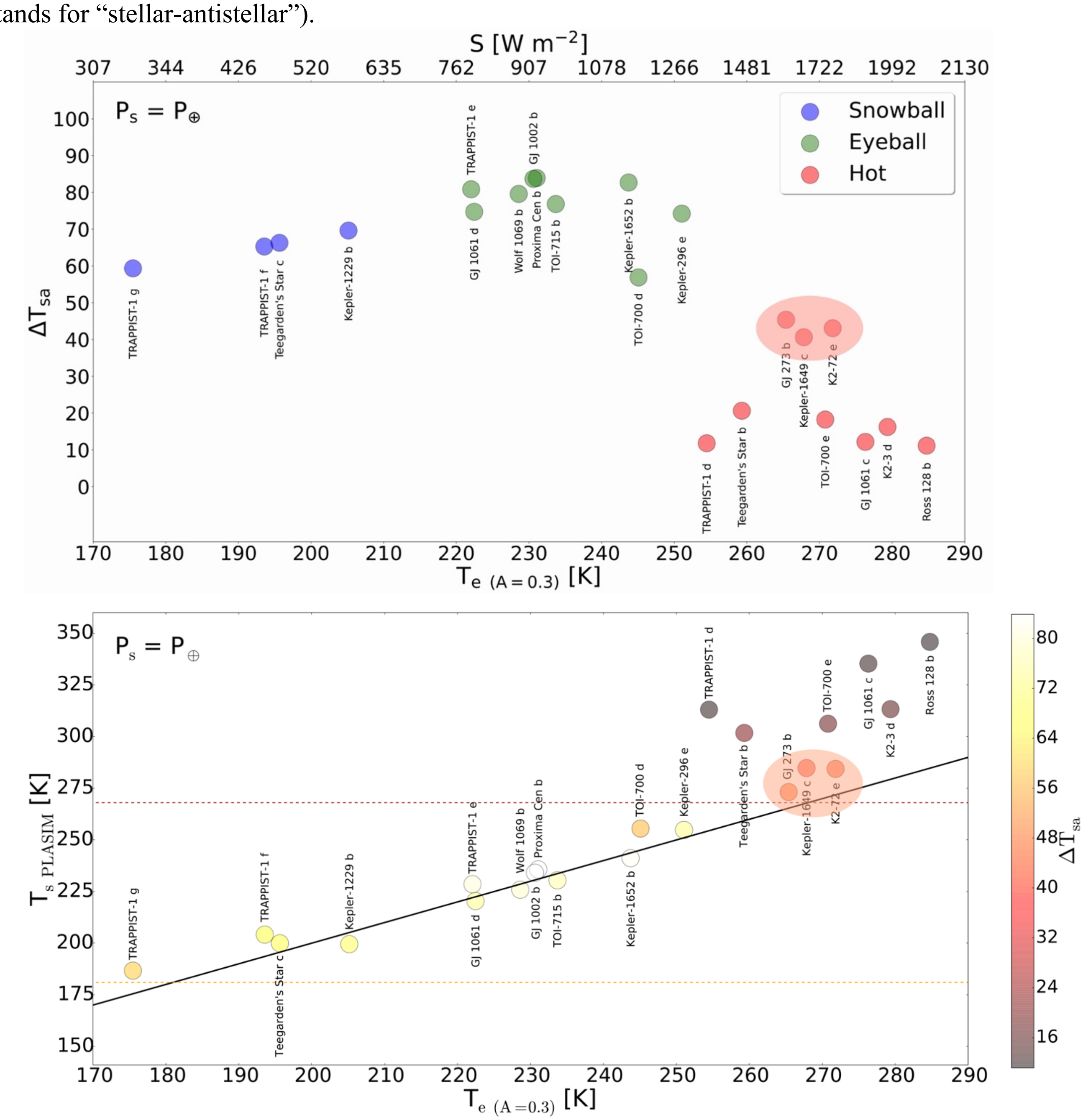

**Figure 6**. Classification of exoplanets into the three climate regimes identified in Figure 2 and discussed

in the main text. In both panels, the lower abscissa is $T_{e\ (A\ =\ 0.3)}$, the emission temperature assuming a terrestrial albedo ($A_{Bond}$ = 0.30). Upper panel: difference between substellar and antistellar point temperatures ($\Delta T_{sa}$, Eq. 5) versus $T_{e\ (A\ =\ 0.3)}$; upper abscissa reports the instellation. Lower panel: surface average temperature calculated by `PLASIM` versus $T_{e\ (A\ =\ 0.3)}$; the color map indicates the $\Delta T_{sa}$ values. The horizontal dashed lines demarcate the transitions between the identified three regimes: $T_{s\ \mathrm{PLASIM}}$ > 270 K; 180 K < $T_{s\ \mathrm{PLASIM}}$ < 270 K; $T_{s\ \mathrm{PLASIM}}$ < 180 K. The diagonal line represents the condition $T_{s\ \mathrm{PLASIM}} = T_{e\ (A\ =\ 0.3)}$. The semi-transparent ellipses refer to the Transitional outliers.

The two panels of Figure 6 further explore the classification between the three regimes introduced above, in terms of $\Delta T_{sa}$. In the upper panel, we plot the $\Delta T_{sa}$ values versus the emission temperature and assumed a terrestrial albedo ($A_{Bond}$ = 0.30). We chose to estimate the planetary emission temperature by using Earth's albedo, rather than the albedo computed by `PLASIM`, in order to derive the parameter directly from an observed quantity, the stellar flux, without introducing additional assumptions about the planet's characteristics. The choice of the emission (blackbody) temperature was made to allow for an immediate comparison with that of Earth (≈ 255 K). For example, Hot planets exhibit emission temperatures comparable or slightly higher than Earth's, despite having significantly higher surface temperatures according to our calculations. For the purpose of clarity, we also display the instellation in the upper x-axis. As for the pressure, we assumed the terrestrial pressure value after having tested that the lower and upper pressures did not qualitatively alter the results (as shown in Figure 3, the main impact is a change in the temperatures).

As shown in the upper panel of Figure 6, three different climate regimes can be identified. Planets marked in blue correspond to Snowball states, with surface temperatures that never exceed the freezing point of water, $T_{freeze}$. This occurs because these planets are located too far from their host star and thus receive insufficient stellar flux to prevent a permanent Snowball state. In red, we marked the planets that are, instead, quite hot: these planets have a surface temperature that is almost everywhere higher than $T_{freeze}$. The remaining planets (shown in green) are characterized by an Eyeball planet temperature structure, where only their illuminated side exhibits a region where the temperature is higher than $T_{freeze}$, with the rest of the surface frozen. These two groups are separated by their $\Delta T_{sa}$ values; the temperature difference was always smaller than 30 K (except for the three outliers GJ 273 b, Kepler-1649 c, and K2-72 e—formerly identified as Transitional and highlighted here by a transparent ellipse (cf. Section 4.1)—that exhibit a temperature difference near 40 K) for Hot planets and higher than 30 K for the Eyeball planets. Thus, aside from the planets in the Transitional regime, the heat distribution in Hot planets was found to be more uniform than that of Eyeball planets. This result can also be inferred by visual inspection of Figure 2. We tested the possibility that changing the definition of $\Delta T$ (for example, by using the average temperature on the dark and illuminated sides of the planets) did not qualitatively alter such findings.

The lower panel of Figure 6 shows the average surface temperature as calculated by `PLASIM` ($T_{s\ \mathrm{PLASIM}}$, see Figure 2) versus $T_{e\ (A\ =\ 0.3)}$ and $\Delta T_{sa}$ (the latter expressed by the color scale). As above, the pressure was set to 1 $P_{\oplus}$. Again, we tested that the lower and upper pressure values did not qualitatively alter the result. Since in our case changing the pressure also meant changing the $CO_2$ mass, we can conclude that altering the greenhouse gases inventory by a factor of 10 did not affect our classification. Note that while Snowball and Eyeball planets followed a similar linear trend in the $\Delta T_{sa}$ versus $T_{s\ \mathrm{PLASIM}}$ plane, non-Transitional Hot planets were consistently offset by approximately 40 K. Despite this finding, the heat distribution and computed surface temperature remained comparable for Snowball and Eyeball regimes, with the primary distinction being the total amount of radiation received.

Besides confirming previous results—namely, the existence of three distinct climate regimes, color-coded in the figure—this plot provides the following additional insights:

i. According to our model, the three regimes are defined by the following surface temperature

thresholds: Hot planets ($T_s > 270$ K), Eyeball planets (180 K < $T_s$ < 270 K), and Snowball planets ($T_s < 180$ K).

ii. Hot planets behaved as expected: their greenhouse effect elevates the surface temperature, while efficient atmospheric heat transport maintains the dark-side temperature above $T_{freeze}$. In contrast, the situation for Eyeball planets is markedly different. In this regime, heat transport was insufficient to prevent ice accumulation on the dark side and across portions of the illuminated hemisphere. Consequently, these planets exhibited a more pronounced day-night temperature contrast, a characteristic typical of low-temperature, tidally locked worlds.

iii. We identified a second key distinction between the climate regimes on Hot and Eyeball planets. Hot planets exhibited average surface temperatures 10–60 K higher than their emission temperatures, whereas the average $T_s$ of Eyeballs remained close to the equilibrium temperature (computed assuming an Earth-like albedo). This discrepancy further highlights the disparate efficiency of interhemispheric heat transports between the two groups.

Qualitatively similar results—particularly regarding the separation in average temperatures and $\Delta T_{sa}$—were obtained for surface pressures of 0.5 $P_\oplus$ and 5 $P_\oplus$.

### *4.6. Habitability*

In this section, we quantify habitability according to two temperature-based indices: the liquid water habitability index, $h_{lw}$, with temperature range between the freezing and boiling points of water, and the biologically motivated index, $h_{50}$, introduced by Silva et al. (2017a). This index adopts a narrower temperature range, 0 °C < $T$ < 50 °C, identified as suitable to represent conditions conducive to sustained, metabolically active biospheres that could plausibly influence atmospheric composition and generate detectable biosignatures. In both cases, the indices were computed as the fraction of the planetary surface whose temperature falls within the specified bounds. This choice was motivated by our aquaplanet configuration, in which water availability is not a limiting factor and temperature provides a first-order diagnostic of surface habitability. By contrast, studies that include continents often employ more complex metrics that combine temperature with hydrological diagnostics (e.g., precipitation, evaporation, or evapotranspiration) to quantify water availability also on land (Del Genio et al., 2019). While our primary analysis established this thermal baseline, an examination of additional surface climate variables, which included precipitation and evaporation, is presented in Section 4.8.

The upper panel of Figure 7 shows the habitability of exoplanets as a function of their surface temperature as calculated by `PLASIM` (on the x-axis), radius (proportional to the size of the circles), and $\Delta T_{sa}$ (the difference between substellar and antistellar temperatures), which serves as a proxy for the prominence of the Eyeball climate regime, as indicated by the color scale. According to our definition and consistent with Figure 6, 18 of the 22 planets are habitable: 9 of them are Hot, and 9 are Eyeballs (for terrestrial configuration); the remaining 4 are frozen.

The separation between the two groups (Eyeball versus Hot planets) at about 270 K is distinctive. For 3 out of the 9 hottest planets, $h_{50}$ is significantly lower than $h_{lw}$. It is conceivable that these planets may experience substantial water vapor buildup, potentially leading to water loss via hydrogen escape. Indeed, Figure 4 demonstrates that Hot planets exhibit significant amounts of water vapor in the upper atmosphere. However, assessing these effects would require photochemical modeling, which is not currently included in `PLASIM`. With the exception of Transitional cases, the two habitable planetary groups remain well separated in terms of their surface temperature.

An alternative representation of the previous result is provided in the upper panel of Figure 8. In this plot, habitability is shown as a function of $T_{e\,(A=0.3)}$, which represents the blackbody emission temperature calculated assuming a terrestrial albedo. All other parameters remained consistent with those used in Figure

7. The emerging picture is consistent with the findings shown in Figure 7. However, as noted above, $T_e$ offers the advantage of being directly derivable from observations. It is therefore crucial to highlight that the distinction between Eyeball and Hot planets persists when using an observationally derived quantity, rather than relying solely on numerical model outputs.

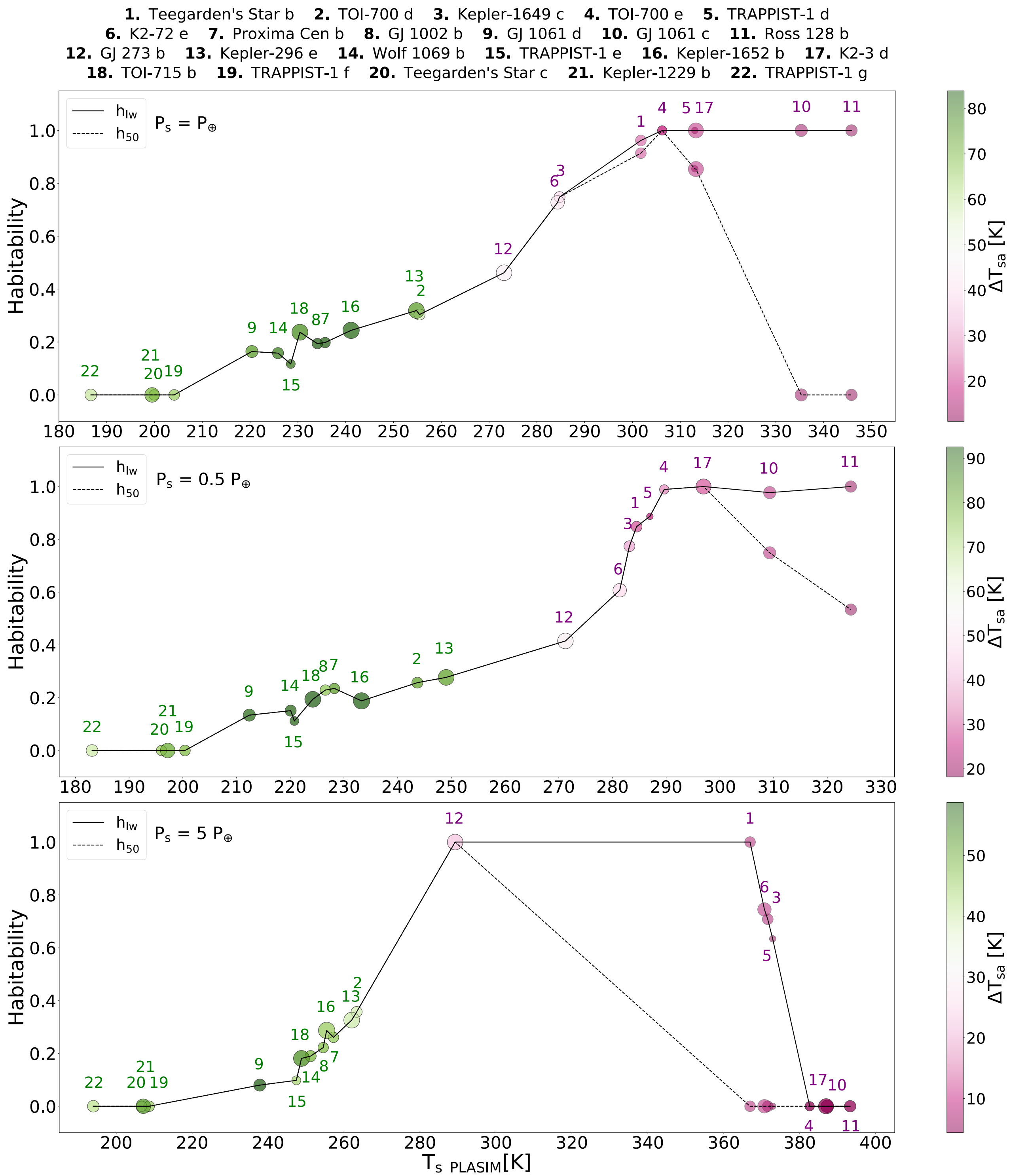

**Figure 7**. Habitability of exoplanets as a function of their surface temperature as calculated by PLASIM

(on the x-axis), radius (rescaled as the size of the circles), and $\Delta T_{sa}$ (difference between substellar and antistellar temperatures, as indicated by the color map). Upper plot: $P_s$ = 1 bar; middle plot: $P_s$ = 0.5 bar; lower plot: $P_s$ = 5 bar. Regarding the habitability definition, we adopted two distinct criteria: the range [$T_{freeze} - T_{boil}$], where the thresholds correspond to the freezing and boiling points of water at the given pressure ($h_{lw}$; solid line), and the fixed range [273.15 − 323.15] K ($h_{50}$; dashed line). The habitability values represent the fraction of the planetary surface where temperatures fall within these defined ranges.

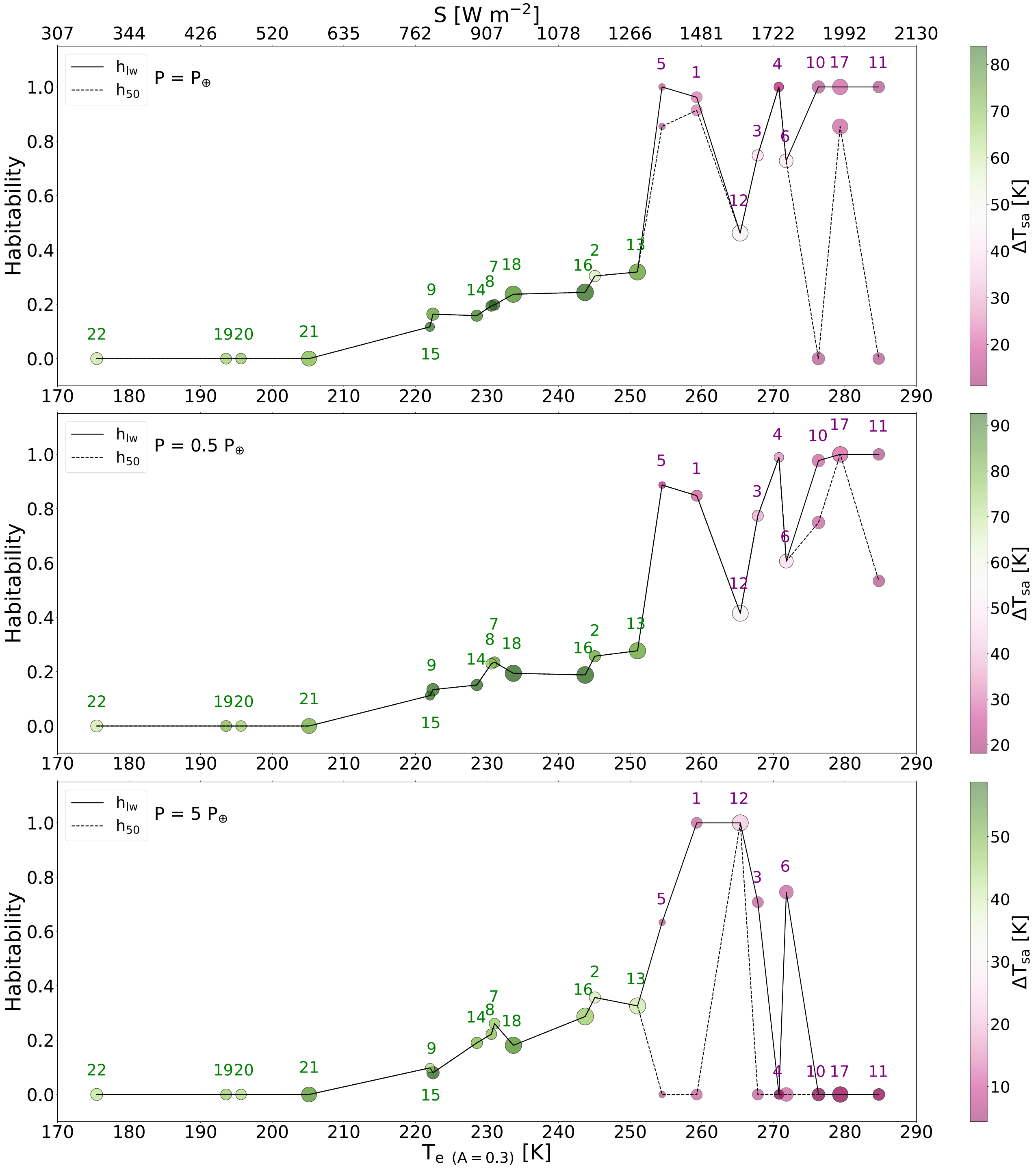

**Figure 8**. Habitability of the simulated exoplanets, as in Figure 7, but represented as a function of their emission temperature $T_{e\ (A\,=\,0.3)}$ (lower abscissa) and instellation (upper abscissa).

Furthermore, we investigated potential dependencies between orbital/rotational periods and habitability indices; however, no statistically significant correlations were identified. Indeed, as illustrated in Figure 4, rotational dynamics appeared to exert only a secondary influence on the global climate state, with the magnitude of the incident stellar flux remaining the primary driver of surface habitability.

### 4.7. Effect of pressure on habitability

The middle and lower panels of Figure 7 and Figure 8 illustrate the habitability of the 22 exoplanets as a function of their surface temperature (as calculated by PLASIM) and of their emission temperature, respectively, for the two different pressures $P_s = 0.5\ P_\oplus$ and $P_s = 5\ P_\oplus$. All other parameters remain unchanged. Consistent with the previously discussed dependence of surface temperature on atmospheric pressure, the shifts in the $T_s$-habitability and $T_e$-habitability relations followed the expected trends. Specifically, average surface temperatures decreased at lower pressures. However, the habitability fractions of Eyeball planets did not change significantly with variations in surface pressure.

At $P_s = P_\oplus$, four Hot planets exhibited $h_{lw} < 1$, indicating that portions of their surface remained below the freezing point ($T_{freeze}$). At the lower pressure of $P_s = 0.5\ P_\oplus$, Hot planets benefited from an enhanced $h_{50}$, which indicates a larger habitability area; this suggests they may also be less prone to water vapor accumulation, which could otherwise lead to moist or runaway greenhouse conditions. Conversely, at $P_s = 5\ P_\oplus$, eight out of nine Hot planets exhibited $h_{50} = 0$, which signifies that their entire surface temperature exceeded 50 °C. In these cases, substantial water vapor accumulation would become increasingly likely, potentially approaching the onset of a moist or runaway greenhouse regime. In contrast, Eyeball worlds remained qualitatively unaffected by the choice of habitability metric, with Figure 8 showing similar trends. In conclusion, a key finding of this study is that the classification of simulated planets into three distinct climatic regimes remains robust and independent of surface pressure within the explored range.

### 4.8. Habitability and the hydrological cycle

To further probe the sensitivity of our habitability assessment, we applied hydrological constraints to a global aquaplanet. By deliberately disregarding the global shallow ocean as a valid freshwater source, we treated the planet as if it were “functionally arid” despite its composition. This approach allowed us to isolate the role of the active hydrological cycle as the sole provider of life-sustaining water. While this represents a provocative and arguably counter-intuitive constraint for an ocean world, it serves as a rigorous stress test for the habitability envelope, identifying regions where atmospheric transport alone is sufficient to maintain a biosphere, independent of the underlying reservoir.

To this end, we adopted the habitability index proposed by Woodward et al. (2025), defined as

$$H\ (T_S, P_R, E) = \begin{cases} 1 & \text{if } 0\ °\text{C} < T_S < 50\ °\text{C}, \quad P_R - E > 0, \quad \text{and } P_R > 250\ \text{mm yr}^{-1} \\ & 0 \text{ otherwise} \end{cases} \qquad (6)$$

where $T_s$ denotes the surface temperature, $P_R$ the precipitation, and $E$ the evaporation. This criterion was applied to each grid cell of the model, and the planetary habitability $h_{p250}$ (indicating habitability with annual precipitation exceeding 250 mm yr$^{-1}$) was subsequently calculated as the area-weighted mean of $H$.

Figure 9 presents a comparison between the $h_{50}$ and $h_{p250}$ indices. Generally, planets with a non-zero $h_{50}$ also maintain a non-zero $h_{p250}$, as surface temperature remains the primary constraint, falling below 273.15 K for Snowball states and exceeding 323.15 K in the two hottest cases. However, the inclusion of the hydrological cycle reduced the estimated habitable area, particularly for Hot planets. This reduction was not driven by a scarcity of precipitation; indeed, rainfall consistently exceeded the 250 mm $yr^{-1}$ threshold on Eyeball planets and remained well above this limit across the majority of the surface of Hot worlds. Instead, the limitation arose in regions where evaporation exceeded precipitation ($P_R - E < 0$), even where the latter was substantial. This phenomenon is observed at the periphery of the Eye on Eyeball planets and characterizes large portions of the surface on Hot planets—most notably at and near the substellar point. We speculate that this behavior is independent of the presence of a global ocean, as a limited water supply would likely exacerbate the deficit. However, a quantitative assessment of more complex habitability metrics will require simulations that incorporate both oceans and continents.

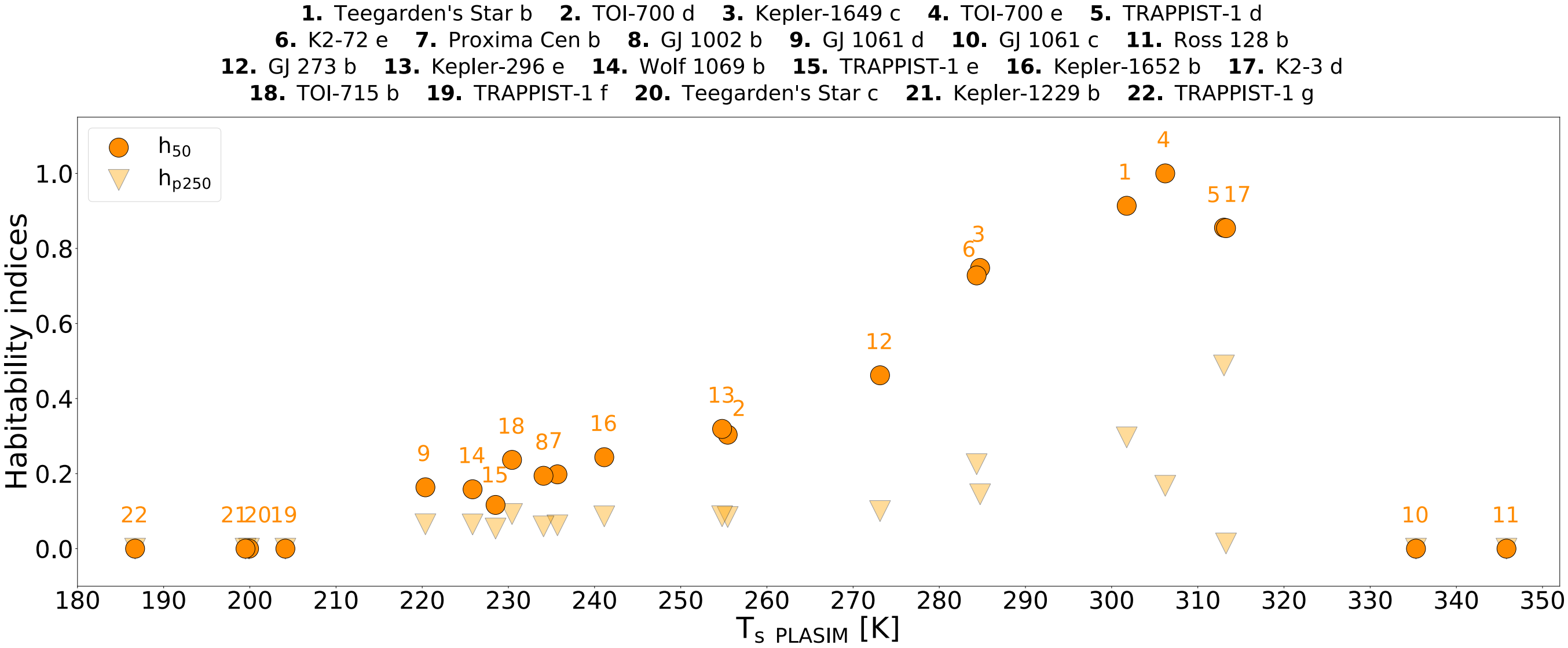

**Figure 9**. Comparison of habitability indices calculated with and without the inclusion of hydrological constraints (precipitation and evaporation) affecting the freshwater availability required for complex life (Woodward et al., 2025).

## 5. Discussion

### 5.1. Framework novelty and climatic regime classification

Our analysis of the 22 observed exoplanets is consistent with previous studies that employed 3D climate models, which highlight the complexity of simulating the climates of tidally locked planets and assessing their habitability under varying atmospheric and orbital conditions. The main novelty of the present study is the consistent investigation of observed targets using the same theoretical framework, adopting the simplest possible configuration.

The planetary sample analyzed in this study enabled the identification of three fundamental climatic regimes: “Eyeballs,” “Hot,” and “Snowballs”—the first two of which are potentially habitable. We found that Eyeball planets exhibit a substellar region with surface temperatures T > 273.15 K, where clouds and a hydrological cycle are mostly confined, as shown in Figure 4. Although we strove to resolve the vertical structure of the atmosphere as accurately as possible, it is important to note that GCMs—including full-scale models used for studying Earth’s climate variability—employ primitive equations to model dynamics. This typically relies on the hypothesis of hydrostatic equilibrium and almost always involves the parametrization of key physical processes, such as convection. These parametrizations are tuned to Earth’s atmospheric circulation. However, tidally locked worlds operate within a fundamentally different physical regime, as their atmospheric dynamics are usually dominated by strong convection at the substellar point. This aspect remains an open question that deserves further study.

### *5.2. Dominance of stellar flux and day-night temperature contrast*

Our computations revealed that stellar flux remains the dominant factor influencing habitability; it outweighs both planetary characteristics and rotational regimes. In particular, we found that instellation—expressed here as the emission temperature, a quantity that is observationally estimated—plays a crucial role in distinguishing Eyeball from Hot planets. In the latter case, atmospheric heat transport was efficient enough to eliminate global ice cover, significantly reducing temperature contrast between the day and night hemispheres.

We identified a threshold $T_e$ between 250 and 260 K that separates the two regimes, a boundary that appears remarkably independent of atmospheric pressure within our simplified framework (Figure 6). This transition is clearly reflected in Figure 5, where the day-night temperature contrast serves as the primary metric to distinguish Snowball and Eyeball worlds from Hot ones. Our finding that this contrast is significantly reduced in the Hot regime is consistent with the results of Haqq-Misra et al. (2018). Interestingly, within the Hot class, we identified three planets occupying a "Transitional" regime—exhibiting intermediate climatic properties between completely ice-free and Eyeball states. Whether this behavior warrants a distinct classification depends upon the extent of the transition relative to stellar flux; this remains a subject for further investigation to determine the statistical likelihood of planets occupying such intermediate states.

### *5.3. Atmospheric pressure effects and high-temperature habitability limits*

For several Hot planets, we found $h_{50} < h_{lw}$, which indicates that a significant portion of the surface lies within the temperature range between 50 °C and the pressure-dependent boiling point of water (Figure 7 and Figure 8). This suggests the possibility of extensive evaporation and elevated water vapor concentrations in the atmosphere, potentially approaching conditions associated with a moist or runaway greenhouse regime.

For clear-sky present-day Earth atmospheres, Kopparapu et al. (2013) provided the temperature thresholds required to trigger a runaway greenhouse state, as a function of stellar effective temperature and orbital distance. According to their results, and with the exception of TRAPPIST-1 d and Teegarden's Star b, all the Hot planets in our sample set would fall within the runaway greenhouse regime. Clearly, this possibility is highly sensitive to atmospheric parameters: in our computations, halving the surface pressure and thus the $CO_2$ column mass made many of these worlds habitable or at least less prone to runaway greenhouse conditions. At a pressure of 1 bar, Ross 128 b was almost certainly within this regime, with an average surface temperature of about 345 K and a maximum temperature of approximately 353 K. Its atmosphere was completely saturated with water vapor at all pressures, indicating that `PLASIM` simulations are likely not correctly capturing the climate dynamics. GJ 1061 c also exhibited a high average surface temperature of ~335 K and a maximum of ~347 K; however, `PLASIM` computations did not show a fully saturated atmosphere. Instead, the maximum specific humidity here was on the order of 0.1 kg $kg^{-1}$, below and east of the substellar point at σ values as low as 0.03, corresponding to an altitude of more than 70 km. No other simulation showed a similar level of specific humidity at such high altitudes.

Here we simply point out that a runaway greenhouse scenario may be likely for planets belonging to the Hot class. A detailed assessment of whether a moist or runaway greenhouse effect may occur on these Hot planets would require detailed, planet-specific studies. Conversely, Eyeball planets maintained their habitability fraction regardless of whether $h_{50}$ or $h_{100}$ was used.

### 5.4. ESI versus climatic habitability

Snowball planets are included in the HWC, as they meet the criteria of being likely rocky planets potentially capable of supporting surface liquid water. While these planets exhibit a relatively high ESI, it is important to note that the Earth Similarity Index and habitability are complementary but not necessarily correlated metrics (Schulze-Makuch et al., 2011). Indeed, a planet can achieve a high ESI score primarily due to its radius and stellar flux matching Earth's, even if its actual climatic conditions are inhospitable. Our simulations confirmed that liquid water is absent from the surface of these Snowball states, even when considering a dense atmosphere with a $p_{CO2}$ five times higher than that of modern Earth.

These findings collectively reveal a significant asymmetry in planetary climate sensitivity: while the habitability of Hot planets is critically dependent on atmospheric composition and surface pressure, we found that increasing the surface pressure of Cold worlds by a factor of five failed to prevent a Snowball climatic state. This suggests that once a planet enters a deep glaciation regime, the atmospheric buffering required to escape it is substantially greater than that needed to stabilize a world on the edge of a runaway greenhouse.

### 5.5. Surface albedo and the aquaplanet

It should be noted that our treatment of surface albedo represents a necessary first approximation. We recomputed the albedo for water, snow, and sea ice for different stellar spectra, following the dual short-wave band approach of Paradise et al. (2022) (cf. Section 3.3). Specifically, for planets orbiting M dwarfs, the spectral sensitivity of ice and snow albedo became a dominant climatic driver. As noted by Shields et al. (2013), the increased absorption of near-infrared radiation by icy surfaces around cooler stars can significantly weaken the ice-albedo feedback, potentially preventing the global glaciation states that would otherwise occur around solar-type stars.

Furthermore, although the present study focused on an aquaplanet configuration, the transition to land-dominated scenarios would introduce additional complexities. Rushby et al. (2020) demonstrated that the mineral composition of continental landmasses, through its interaction with the host star's spectral energy distribution, can shift global mean surface temperatures by as much as 50 K. Consequently, while our approach captured the primary energy balance of a water-dominated world, incorporating surface mineralogy will be essential for future high-fidelity simulations of exoplanetary climates.

### 5.6. Climate bistability and sensitivity to atmospheric mass

We found no evidence of climate bistability—defined here as the existence of two distinct stable climatic equilibria, such as Snowball and temperate states on Earth, when varying the initial temperature. Specifically, repeating our simulations with a "cold start" ($T = 220$ K) yielded the same equilibrium states as the "hot start" runs, a result consistent with the findings of Checlair et al. (2019).

While other forms of climate bistability are possible—such as the sensitivity to $CO_2$ forcing described by Wolf et al. (2018), where doubling $CO_2$ content can shift average surface temperatures by as much as 21 K—our results showed a different but equally significant sensitivity to atmospheric mass. By varying the surface pressure (and thus the $CO_2$ mass) between 0.5 and 5 bar, we observed global average temperature shifts of up to 93 K in the most extreme case, TOI-700 d. When we grouped the planets into Snowball, Eyeball, and Hot classes, the average temperature changes across this pressure range were 9.9, 24.3, and 84.5 K, respectively. Notably, despite these substantial shifts in absolute surface temperature, the overall temperature distribution—and consequently the distinction between the three classes—remained robust. A more detailed investigation into how temperature distributions respond to varying greenhouse gas concentrations is, however, beyond the scope of the present study.

### *5.7. Comparison with existing literature*

Among recent studies that have investigated the climate of tidally locked planets orbiting M dwarfs using simplified 3D models, we highlighted in Section 3 the work of Paradise et al. (2022), who adapted `PLASIM` for such systems. Our results are in agreement with their findings. For instance, the surface temperature map of TRAPPIST-1 e, presented in Figure 2, is in excellent agreement with their Figure 21. In particular, both simulations identified cold stripes at the north and south poles ($T \approx 175$ K), as well as regions of upwelling and downwelling (near the north and south poles, respectively) characterized by similar temperature profiles.

The `THAI` project, which aimed to compare four different 3D climate models, found general qualitative agreement among the simulations despite significant variability in certain cases. Specifically, Sergeev et al. (2022) investigated the TRAPPIST-1 system in the "moist" scenario, assuming an aquaplanet configuration. Our simulation of TRAPPIST-1 e aligns closely with theirs, particularly with the results from the `ROCKE-3D` model, exhibiting only minor discrepancies near the substellar point. However, we note that our simulation yielded a slightly cooler climate, with a mean temperature of approximately 229 K, a minimum of 178 K, and a maximum of 284 K. These values are comparable to those produced by the `Unified Model (UM)` in the same study. Consequently, many of the conclusions drawn by the `THAI` project are directly applicable to our findings as well.

Galuzzo et al. (2021) investigated Proxima Centauri b using `PLASIM`, which they also adapted for tidally locked planets orbiting M dwarfs. Assuming a modern Earth-like atmosphere and an aquaplanet configuration, their simulated surface temperature maps are qualitatively consistent with our Figure 2. Similarly, Lobo and Shields (2024) simulated tidally locked Earth-like candidates orbiting the M3.0V star AD Leo—which currently lacks confirmed exoplanets (Kossakowski et al., 2022, and references therein)—and the K2 V star $\epsilon$ Eri. By varying stellar instellation, and consequently orbital and rotational periods, they focused on slow rotators. In contrast, our sample set included slow, fast, and intermediate rotators. This distinction is crucial: slow rotators are dominated by heat transport driven by the day-night temperature gradient centered on the substellar point, whereas fast rotators exhibit strong latitudinal (zonal) winds, and intermediate cases display characteristics of both regimes.

Lobo and Shields (2024) identified two habitability types: Eyeball and Terminator-habitable states. Since the latter emerges only in land-planet configurations, we could not confirm its existence within our aquaplanet framework. Nevertheless, our Eyeball planets exhibited similar behavior to theirs; specifically, precipitation patterns were largely, though not entirely, confined to the substellar region (see Figure 4). As in their simulations, water could become trapped on the night-side hemisphere. However, as noted by Lobo and Shields (2024), the physical mechanisms for returning water to the day side, such as ice melting at or near the terminator, remain poorly understood. In a true aquaplanet, this issue is less critical unless the entire ocean freezes, making the phenomenon more pertinent to land-dominated worlds.

### *5.8. Habitability metrics*

Several habitability metrics have been proposed that leverage spatially resolved combinations of variables derived from 3D models. These metrics often couple temperature with hydrological diagnostics, such as precipitation and evaporation, to better capture the multifaceted nature of habitable environments (e.g., Stevenson 2019; Del Genio et al., 2019; Adams et al., 2025; Woodward, Rushby and Mayne 2025). In this work, however, we intentionally adopted a very simple configuration—an aquaplanet with a 1 bar surface pressure and synchronous rotation—to isolate the influence of fundamental planetary and astrophysical parameters. Given the absence of land-based water supply limitations in our model, near-surface humidity remains primarily governed by surface temperature (see Figure 4). Consequently, any habitability index based on humidity would likely yield little independent information beyond that already captured by temperature-based indices. Nevertheless, in Section 4.8 we examined the influence of the hydrological cycle on habitability under the specific assumption that the global ocean does not function as a direct water source for the biosphere. Our results demonstrated that intense evaporation can indeed act as a significant limiting factor. Furthermore, while introducing nutrient limitations would not be meaningful in this idealized configuration, it would be valuable to investigate limitations imposed by illumination, particularly for photosynthetic organisms (Maris et al., in preparation).

## 6. Summary and Conclusions

We investigated the habitability of 22 observed tidally locked exoplanets orbiting M dwarf stars, selected from those currently exhibiting the largest Earth-similarity index (ESI) values. These terrestrial planets represent some of the most promising targets in the ongoing and future search for habitable worlds beyond our solar system.

We evaluated their potential climates using consistent assumptions across a range of planetary and atmospheric parameters, specifically exploring surface pressures of $P_s$ = 0.5, 1, and 5 $P_\oplus$ with a fixed $CO_2$ mixing ratio of 360 ppm. Climate simulations were performed using our modified version of the intermediate-complexity 3D climate model `PLASIM-LSG`. Habitability was quantified via two temperature-based metrics: the liquid water habitability index, $h_{lw}$, and the biologically motivated index, $h_{50}$ (Silva et al., 2017a).

Within the scope of our simplified model and the parameter range explored, our analysis revealed three distinct climate regimes:

- Snowball planets: TRAPPIST-1 f, Teegarden's Star c, Kepler-1229 b, TRAPPIST-1 g. These planets are globally frozen and effectively uninhabitable.
- Eyeball planets: TOI-700 d, Proxima Cen b, GJ 1002 b, GJ 1061 d, Kepler-296 e, Wolf 1069 b, TRAPPIST-1 e, Kepler-1652 b, TOI-715 b. These worlds maintain habitability relatively independently of atmospheric pressure, within the limits of the modern-Earth atmospheric composition adopted here.
- Hot planets: Teegarden's Star b, Kepler-1649 c, TOI-700 e, TRAPPIST-1 d, K2-72 e, GJ 1061 c, Ross 128 b, GJ 273 b, K2-3 d. These planets may achieve habitability only under specific atmospheric conditions that facilitate sufficient cooling. Notably, three of these (GJ 273 b, Kepler 1649 c, and K2-72 e) have an only partially melted night hemisphere, representing a transitional regime between the Hot and Eyeball classes.

While these planets were selected for their high Earth Similarity Index (ESI) values, we found that ESI and habitability are not necessarily correlated (consistent with Schulze-Makuch et al., 2011). This lack of correlation reinforces the necessity of dedicated climate modeling; in particular, Eyeball planets—arguably the most promising candidates for habitability—are distributed across a wide range of ESI values.

At lower atmospheric pressures, we found that the average surface temperatures decreased. While Eyeball planets remained largely unaffected and maintained stable habitability, Hot planets benefited from improved habitability (as measured by $h_{50}$), which rendered them less susceptible to water loss or greenhouse instabilities. For several of these planets, $h_{lw} < 1$, which indicates that portions of their surface remained below the freezing point of the water. At higher pressures, this behavior reverses: Eyeball planets still maintained stable climates, but the habitability of Hot planets was reduced significantly in terms of $h_{50}$, a condition that may favor the onset of greenhouse-driven instabilities. Despite these pressure-dependent effects, the overall classification into Snowball, Eyeball, and Hot regimes—and their distinct thermal characteristics—remained robust across the explored pressure range. We note that since the $CO_2$ mixing ratio was kept constant, varying the surface pressure in our setup also inherently modified the total greenhouse gas mass, thereby strengthening the greenhouse effect.

These results suggest that planets with emission temperatures $T_e < 255$ K (corresponding to an instellation $S < 1460$ W m$^{-2}$) should be prioritized as the most promising candidates for habitability and future biosignature surveys. Such worlds can sustain liquid surface water across a broader range of atmospheric conditions and are inherently more resilient to climatic instabilities. This is consistent with the findings of Lobo and Shields (2024), although we further argue that even Hot planets may retain habitability under specific atmospheric configurations. Notably, this $T_e$ threshold aligns closely with the inner edge of the habitable zone for M dwarf stars, as calculated by, for example, Kopparapu et al. (2013).

In future work, we plan to explore systematically a broader range of planetary configurations, including diverse atmospheric compositions—specifically varying greenhouse gas strengths—as well as different surface geographies and ocean models. This ongoing effort aims to provide theoretical support and inform target selection for future space missions dedicated to the detection and characterization of potentially habitable exoplanets.

## Acknowledgements

This study is supported by the Italian Space Agency (ASI) within the `ASTERIA` project (ASI N. 2023-5-U.0). We acknowledge financial support from the Agencia Estatal de Investigación (AEI/10.13039/501100011033) of the Ministerio de Ciencia e Innovación, and the ERDF "`A way of making Europe`" through project PID2022-137241NB-C42. Part of this work has also been supported by the INAF RSN2 `ClimHAB-RBA` (`Climate Habitability: Refraction and Biosignatures in Habitable Exoplanet Atmospheres`) MINI-GRANT N. 1.05.12.04.02. We sincerely thank the anonymous referees for the insightful comments and constructive feedback provided.

## References

Adams AD, Colose C, Merrelli A, et al. Habitability in 4D: Predicting the climates of Earth analogs across rotation and orbital configurations. *Astrophys J* 2025;981(1):98–12; https://doi.org/10.3847/1538-4357/ada3c8

Agol E, Dorn C, Grimm SL, et al. Refining the transit-timing and photometric analysis of TRAPPIST-1: Masses, radii, densities, dynamics, and ephemerides. *Planet Sci J* 2021;2(1):1; https://doi.org/10.3847/PSJ/abd022

Angeloni M, Palazzi E, von Hardenberg J. Evaluation and climate sensitivity of the PlaSim v.17 Earth System Model coupled with ocean model components of different complexity. 2020. Available from https://gmd.copernicus.org/preprints/gmd-2020-245 [Last accessed: August 15, 2025].

Astudillo-Defru N, Forveille T, Bonfils X, et al. The HARPS search for southern extra-solar planets. XLI. A dozen planets around the M dwarfs GJ 3138, GJ 3323, GJ 273, GJ 628, and GJ 3293. *Astron Astrophys* 2017; 602:A88; https://doi.org/10.1051/0004-6361/201630153

Barclay T, Quintana EV, Adams FC, et al. The five planets in the Kepler-296 binary system all orbit the primary: A statistical and analytical analysis. *Astrophys J* 2015, 809(1):7–16; https://doi.org/10.1088/0004-637X/809/1/7

Barnes R. Tidal locking of habitable exoplanets. *Celest Mech Dyn Astron* 2017;129(4):509-536; https://doi.org/10.1007/s10569-017-9783-7

Barnes R., Meadows VS, Evans N. Comparative habitability of transiting exoplanets. *Astrophys J* 2015;814:91–102; https://doi.org/10.1088/0004-637X/814/2/91

Bochanski JJ, Hawley SL, Covey KR, et al. The luminosity and mass functions of low-mass stars in the galactic disk. II. The field. *Astron J* 2010;139(6):2679–2699; https://doi.org/10.1088/0004-6256/139/6/2679

Bonfils X, Delfosse X, Udry S, et al. The HARPS search for southern extra-solar planets. XXXI. The M-dwarf sample. *Astron Astrophys* 2013;549:A109; https://doi.org/10.1051/0004-6361/201014704

Bonfils X, Astudillo-Defru N, Díaz R, et al. A temperate exo-Earth around a quiet M dwarf at 3.4 parsec. *Astron Astrophys* 2018;613:A25; https://doi.org/10.1051/0004-6361/201731973

Borucki WJ, Agol E, Fressin F, et al. Kepler-62: A five-planet system with planets of 1.4 and 1.6 Earth radii in the habitable zone. *Science* 2013;340(6132):587–590; https://doi.org/10.1126/science.1234702

Checlair JH, Olson SL, Jansen MF, et al. No snowball on habitable tidally locked planets with a dynamic ocean. *Astrophys J Lett* 2019;884(2):L46; https://doi.org/10.3847/2041-8213/ab487d

Chen J, Kipping D. Probabilistic forecasting of the masses and radii of other words. *Astrophys J* 2017;834:17–29; https://doi.org/10.3847/1538-4357/834/1/17

Cifuentes C, Caballero JA, Cortés-Contreras M, et al. CARMENES input catalogue of M dwarfs. V. Luminosities, colours, and spectral energy distributions. *Astron Astrophys* 2020;642:A115; https://doi.org/10.1051/0004-6361/202038295

Cifuentes C. *Astrophysical Parameters of M Dwarfs with Exoplanets*. (PhD thesis) Complutense University of Madrid: Madrid, Spain; 2023.

Del Genio AD, Way MJ, Kiang NY, et al. Climates of warm Earth-like planets. III. Fractional habitability from a water cycle perspective. *Astrophys J* 2019;887(2):197–208; https://doi.org/10.3847/1538-4357/ab57fd

Delrez L, Murray CA, Pozuelos FJ, et al. Two temperate super-Earths transiting a nearby late-type M dwarf. *Astron Astrophys* 2022;667:A59; https://doi.org/10.1051/0004-6361/202244041

Del Genio AD, Suozzo RJ. A comparative study of rapidly and slowly rotating dynamical regimes in a terrestrial general circulation model. *J Atmos Sci* 1987;44:973–986; https://doi.org/10.1175/1520-0469(1987)044<0973:ACSORA>2.0.CO;2

Dole SH. *Habitable Planets for Man*. Blaisdell Publishing Company: Boston, MA; 1964.

Dreizler S, Jeffers SV, Rodríguez E, et al. RedDots: A temperate 1.5 Earth-mass planet candidate in a compact multiterrestrial planet system around GJ 1061. *Mon Not R Astron Soc* 2020;493(1):536-550; https://doi.org/10.1093/mnras/staa248

Dreizler S, Luque R, Ribas I, et al. Teegarden's star revisited. A nearby planetary system with at least three planets. *Astron Astrophys* 2024;684:A117; https://doi.org/10.1051/0004-6361/202348033

Dressing CD, Vanderburg A, Schlieder JE, et al. Characterizing K2 candidate planetary systems orbiting low-mass stars. II. Planetary systems observed during campaigns 1-7. *Astron J* 2017;154(5):207–232; https://doi.org/10.3847/1538-3881/aa89f2

Endl M, Cochran WD, Kürster M, et al. Exploring the frequency of close-in jovian planets around M dwarfs. *Astrophys J* 2006;649(1):436–443; https://doi.org/10.1086/506465

Faria JP, Suárez Mascareño A, Figueira P, et al. A candidate short-period sub-Earth orbiting Proxima Centauri. *Astron Astrophys* 2022;658:A115; https://doi.org/10.1051/0004-6361/202142337

Farnsworth A, Lo Eunice, YT, Valdes PJ, et al. Climate extremes likely to drive land mammal extinction during next supercontinent assembly. *Nat Geosci* 2023;16:901–908; https://doi.org/10.1038/s41561-023-01259-3

Fauchez TJ, Turbet M, Wolf ET, et al. TRAPPIST-1 habitable atmosphere intercomparison (THAI): Motivations and protocol version 1.0. *Geosci Model Dev* 2020;13(2):707–716; https://doi.org/10.5194/gmd-13-707-2020

Fauchez TJ, Villanueva GL, Sergeev DE, et al. The TRAPPIST-1 habitable atmosphere intercomparison

(THAI). III. Simulated observables-the return of the spectrum. *Planet Sci J* 2022;3(9):213–229; https://doi.org/10.3847/PSJ/ac6cf1
Fraedrich K, Jansen H, Kirk E, et al. The planet simulator: Towards a user-friendly model. *Meteorol Z* 2005;14(3):299–304; https://doi.org/10.1127/0941-2948/2005/0043
Fraedrich K. A suite of user-friendly global climate models: hysteresis experiments. *Eur Phys J Plus* 2012;127(5):53–62; https://doi.org/10.1140/epjp/i2012-12053-7
Galuzzo D, Cagnazzo C, Berrilli F, et al. Three-dimensional climate simulations for the detectability of Proxima Centauri b. *Astrophys J* 2021;909(2):191–205; https://doi.org/10.3847/1538-4357/abdeb4
Gilbert EA, Vanderburg A, Rodriguez JE, et al. A second Earth-sized planet in the habitable zone of the M dwarf, TOI-700. *Astrophys J Lett* 2023;944(2):L35; https://doi.org/10.3847/2041-8213/acb599
Grieger B, Segschneider J, Keller HU, et al. Simulating Titan's tropospheric circulation with the portable university model of the atmosphere. *Adv Space Res* 2004;34(8):1650–1654; https://doi.org/10.1016/j.asr.2003.08.079
Hammond M, Lewis NT. The rotational and divergent components of atmospheric circulation on tidally locked planets. *Proc. Natl. Acad. Sci. U.S.A.* 2021;118(13): e2022705118; https://doi.org/10.1073/pnas.2022705118
Haqq-Misra J, Wolf ET, Joshi M, et al. Demarcating circulation regimes of synchronously rotating terrestrial planets within the Habitable Zone. *Astrophys J* 2018;852(2):67–82; https://doi.org/10.3847/1538-4357/aa9f1f
Haqq-Misra J, Wolf E, Joshi M, et al. Erratum: "Demarcating circulation regimes of synchronously rotating terrestrial planets within the Habitable Zone" (2018, ApJ, 852, 67). *Astrophys J* 2020;896:174; https://doi.org/10.3847/1538-4357/ab9a4b
Haqq-Misra J, Wolf ET, Fauchez TJ, et al. The sparse atmospheric model sampling analysis (SAMOSA) intercomparison: Motivations and protocol version 1.0: A CUISINES model intercomparison project. *Planet Sci J* 2022;3(11):260–269; https://doi.org/10.3847/PSJ/ac9479
Hertwig E, Lunkeit F, Fraedrich K. Low-frequency climate variability of an aquaplanet. *Theor Appl Climatol* 2015;121(3–4):459–478; https://doi.org/10.1007/s00704-014-1226-8
Hu Y, Yang J. Role of ocean heat transport in climates of tidally locked exoplanets around M dwarf stars. *Proc Natl Acad Sci USA* 2014;111(2):629–634; https://doi.org/10.1073/pnas.1315215111
Jernigan J, Lafléche É, Burke A, et al. Superhabitability of high-obliquity and high-eccentricity planets. *Astrophys J* 2023;944(2):205–214; https://doi.org/10.3847/1538-4357/acb81c
Joshi MM, Haberle RM, Reynolds RT. Simulations of the atmospheres of synchronously rotating terrestrial planets orbiting M dwarfs: Conditions for atmospheric collapse and the implications for habitability. *Icarus* 1997;129(2):450–465; https://doi.org/10.1006/icar.1997.5793
Kaspi Y, Showman AP. Atmospheric dynamics of terrestrial exoplanets over a wide range of orbital and atmospheric parameters. *ApJ* 2015;804(1):60–77; https://doi.org/10.1088/0004-637X/804/1/60
Kasting JF, Whitmire DP, Reynolds RT. Habitable zones around main sequence stars. *Icarus* 1993;101(1):108–128; https://doi.org/10.1006/icar.1993.1010
Kopparapu RK, Ramirez R, Kasting JF, et al. Habitable zones around main-sequence stars: New estimates. *Astrophys J* 2013;765(2):131–146; https://doi.org/10.1088/0004-637X/765/2/131
Kopparapu RK, Ramirez RM, Schotte|Kotte J, et al. Habitable zones around main-sequence stars: Dependence on planetary mass. *Astrophys J Lett* 2014;787:L29; https://doi.org/10.1088/2041-8205/787/2/L29
Kossakowski D, Kürster M, Henning T, et al. The CARMENES search for exoplanets around M dwarfs. Stable radial-velocity variations at the rotation period of AD Leonis: A test case study of current limitations to treating stellar activity. *Astron Astrophys* 2022;666:A143; https://doi.org/10.1051/0004-6361/202243773
Kuzuhara M, Fukui A, Livingston JH, et al. Gliese 12 b: A temperate Earth-sized planet at 12 pc ideal for atmospheric transmission spectroscopy. *Astrophys J Lett* 2024;967(2):L21; https://doi.org/10.3847/2041-8213/ad3642
Laguë MM, Quetin GR, Ragen S, et al. Continental configuration controls the base-state water vapor greenhouse effect: Lessons from half-land, half-water planets. *Clim Dyn* 2023;61:5309–5330; https://doi.org/10.1007/s00382-023-06857-w
Lobo AH, Shields AL. Climate regimes across the habitable zone: A comparison of synchronous rocky M and K dwarf planets. *Astrophys J* 2024;972(1):71–85; https://doi.org/10.3847/1538-4357/ad58bb
Lunkeit F, Borth H, Böttinger M, et al. *Planet Simulator Reference Manual Version 16*. (Technical Report) Meteorologisches Institut, University of Hamburg: Hamburg, Germany; 2011.
Luque R, Pallé E. Density, not radius, separates rocky and water-rich small planets orbiting M dwarf stars. *Science* 2022;377(6611):1211–1214; https://doi.org/10.1126/science.abl7164
Macdonald E, Paradise A, Menou K, et al. Climate uncertainties caused by unknown land distribution on habitable M-Earths. *Mon Not R Astron Soc* 2022;513(1):2761–2769; https://doi.org/10.1093/mnras/stac1040
Macdonald E, Menou K, Lee C, et al. Climate transition to temperate nightside at high atmosphere mass. *Astrophys J* 2025;981(1):3–17; https://doi.org/10.3847/1538-4357/adb0cb
Maier-Reimer E, Mikolajewicz U, Hasselmann K. Mean circulation of the Hamburg LSG OGCM and its sensitivity to the thermohaline surface forcing. *J Phys Oceanogr* 1993;23(4):731–757; https://doi.org/10.1175/1520-0485(1993)023<0731:MCOTHL>2.0.CO;2
Marfil E, Tabernero HM, Montes D, et al. The CARMENES search for exoplanets around M dwarfs. Stellar atmospheric parameters of target stars with SteParSyn. *Astron Astrophys* 2021;656:A162; https://doi.org/10.1051/0004-6361/202141980
Matsuno, T. Quasi-geostrophic motions in the equatorial area. *J Meteorol Soc Jpn* 1966;Ser. II, 44(1):25–43; https://doi.org/10.2151/jmsj1965.44.1_25
Mayor M, Queloz D. A Jupiter-mass companion to a solar-type star. *Nature* 1995;378(6555):355–359; https://www.scirp.org/reference/referencespapers?referenceid=1661378
Mehling O, Bellomo K, Angeloni M, et al. High-latitude precipitation as a driver of multicentennial variability of the AMOC in a climate model of intermediate complexity. *Clim Dyn* 2023;61(3–4):1519–1534; https://doi.org/10.1007/s00382-022-06640-3
Méndez A, Rivera-Valentín EG, Schulze-Makuch D, et al. Habitability models for astrobiology. *Astrobiology* 2021;21:1017–1027; https://doi.org/10.1089/ast.2020.2342
National Academies of Sciences, Engineering, and Medicine. *Pathways to Discovery in Astronomy and Astrophysics for the 2020s*. National Academies Press: Washington, DC; 2021.
Nelson DL, Cox MM. (eds.) *Lehninger Principles of Biochemistry* (Fourth edition). WH Freeman & Co: New York; 2004.
Noda S. The circulation pattern and day-night heat transport in the atmosphere of a synchronously rotating aquaplanet: Dependence on planetary rotation rate. *Icarus* 2017;282:1–18; https://doi.org/10.1016/j.icarus.2016.09.004
Nutzman P, Charbonneau D. Design considerations for a ground-based transit search for habitable planets orbiting M dwarfs. *Publ Astron Soc Pac* 2008;120(865):317–327; https://doi.org/10.1086/533420
Paradise A, Fan BL, Menou K, et al. Climate diversity in the solar-like habitable zone due to varying background gas pressure. *Icarus* 2021;358:114301; https://doi.org/10.1016/j.icarus.2020.114301
Paradise A, Macdonald E, Menou K, et al. ExoPlaSim: Extending the planet simulator for exoplanets. *Mon Not R Astron Soc* 2022;511(3):3272–3303; https://doi.org/10.1093/mnras/stac172
Peale SJ. Consequences and inferences from tidal interactions in the solar system. *Ann Geophys* 1977;33:23–29; https://ntrs.nasa.gov/citations/19770058162
Pierrehumbert RT. A palette of climates for Gliese 581g. *Astrophys J* 2011;726(1):L8; https://doi.org/10.1088/2041-8205/726/1/L8

Quanz SP, Ottiger M, Fontanet E, et al. Large interferometer for exoplanets (LIFE). I. Improved exoplanet detection yield estimates for a large mid-infrared space-interferometer mission. *Astron Astrophys* 2022;664:A21; https://doi.org/10.1051/0004-6361/202140366

Reylé C, Jardine K, Fouqué P, et al. The 10 parsec sample in the Gaia era. *Astron Astrophys* 2021;650:A201; https://doi.org/10.1051/0004-6361/202140985

Ribas I, Reiners A, Zechmeister M, et al. The CARMENES search for exoplanets around M dwarfs. Guaranteed time observations data release 1 (2016–2020). *Astron Astrophys* 2023;670:A139; https://doi.org/10.1051/0004-6361/202244879

Ricker GR, Winn JN, Vanderspek R, et al. Transiting exoplanet survey satellite (TESS). *J Astron Telescopes Instruments Syst* 2015;1:014003; https://doi.org/10.1117/1.JATIS.1.1.014003

Rushby AJ, Shields AL, Wolf ET, et al. The effect of land albedo on the climate of land-dominated planets in the TRAPPIST-1 system. *Astrophys J* 2020;904(2):124–137; https://doi.org/10.3847/1538-4357/abbe04

Schulze-Makuch D, Méndez A, Fairén AG, et al. A two-tiered approach to assessing the habitability of exoplanets. *Astrobiology* 2011; 11(10);1041–1052; https://doi.org/10.1089/ast.2010.0592

Segschneider J, Grieger B, Keller HU, et al. Response of the intermediate complexity Mars climate simulator to different obliquity angles. *Planet Space Sci* 2005;53(6):659–670; https://doi.org/10.1016/j.pss.2004.10.003

Semtner AJ Jr. A model for the thermodynamic growth of sea ice in numerical investigations of climate. *J Phys Oceanogr* 1976;6:379–389; https://doi.org/10.1175/1520-0485(1976)006<0379:AMFTTG>2.0.CO;2

Sergeev DE, Lambert FH, Mayne NJ, et al. Atmospheric convection plays a key role in the climate of tidally locked terrestrial exoplanets: Insights from high-resolution simulations. *Astrophys J* 2020;894(2):84–102; https://doi.org/10.3847/1538-4357/ab8882

Sergeev DE, Fauchez TJ, Turbet M, et al. The TRAPPIST-1 habitable atmosphere intercomparison (THAI). II. Moist cases-the two waterworlds. *Planet Sci J* 2022;3(9):212–237; https://doi.org/10.3847/PSJ/ac6cf2

Shields AL, Meadows VS, Bitz CM, et al. The effect of host star spectral energy distribution and ice-albedo feedback on the climate of extrasolar planets. *Astrobiology* 2013; 13(8):715–739; https://doi.org/10.1089/ast.2012.0961

Shields AL, Ballard S, Johnson JA. The habitability of planets orbiting M-dwarf stars. *Phys Rep* 2016;663:1–38; https://doi.org/10.1016/j.physrep.2016.10.003

Shields AL. The climates of other worlds: A review of the emerging field of exoplanet climatology. *Astrophys J Suppl Ser* 2019;243(2):30–41; https://doi.org/10.3847/1538-4365/ab2fe7

Showman AP, Wordsworth RD, Merlis TM, Kaspi Y. Atmospheric circulation of terrestrial exoplanets. In: Mackwell SJ, Simon-Miller AA, Harder JW, Bullock MA (eds.), *Comparative Climatology of Terrestrial Planets*. University of Arizona Press; Tucson, AZ; 2014; p. 277–305.

Silva L, Vladilo G, Schulte PM, et al. From climate models to planetary habitability: Temperature constraints for complex life. *Int J Astrobiol* 2017a;16(3):244–265; https://doi.org/10.1017/S1473550416000215

Silva L, Vladilo G, Murante G, et al. Quantitative estimates of the surface habitability of Kepler-452b. *Mon Not R Astron Soc* 2017b;470(2):2270–2282; https://doi.org/10.1093/mnras/stx1396

Spohn T. Exo-geoscience perspectives beyond habitability. *Space Sci Rev* 2026;222;1 (9); https://doi.org/10.1007/s11214-026-01265-y

Spiegel DS, Menou K, Scharf CA. Habitable climates. *Astrophys J* 2008;681(2):1609–1623; https://doi.org/10.1086/588089

Stassun KG, Oelkers RJ, Paegert M, et al. The revised TESS input catalog and candidate target list. *Astron J* 2019;158(4):138–159; https://doi.org/10.3847/1538-3881/ab3467

Stevenson DS. Phytoclimatic mapping of exoplanets. *Int J Astrobiol* 2020;19:68–77;

https://doi.org/10.1017/S1473550419000181
Tarter JC, Backus PR, Mancinelli RL, et al. A reappraisal of the habitability of planets around M dwarf stars. *Astrobiology* 2007;7(1):30–65; https://doi.org/10.1089/ast.2006.0124
Torres G, Kipping DM, Fressin F, et al. Validation of 12 small Kepler transiting planets in the Habitable Zone. *Astrophys J* 2015;800(2):99–122; https://doi.org/10.1088/0004-637X/800/2/99
Torres G, Kane SR, Rowe JF, et al. Validation of small Kepler transiting planet candidates in or near the Habitable Zone. *Astron J* 2017;154(6):264–282; https://doi.org/10.3847/1538-3881/aa984b
Trifonov T, Caballero JA, Morales JC, et al. A nearby transiting rocky exoplanet that is suitable for atmospheric investigation. *Science* 2021;371(6533):1038–1041; https://doi.org/10.1126/science.abd7645
Turbet M, Fauchez TJ, Sergeev DE, et al. The TRAPPIST-1 habitable atmosphere intercomparison (THAI). I. Dry cases—The fellowship of the GCMs. *Planet Sci J* 2022;3(9):211–227; https://doi.org/10.3847/PSJ/ac6cf0
Turner DA, Eschen YNE, Murgas F, et al. The mass of the exo-Venus Gliese 12 b, as revealed by HARPS-N, ESPRESSO, and CARMENES. *Mon Not R Astron Soc* 2026;545(1);1–25; https://doi.org/10.1093/mnras/staf1703
Way MJ, Del Genio AD, Aleinov I, et al. Climates of warm Earth-like planets. I. 3D model aimulations. *Astrophys J Suppl Ser* 2018;239:2–23; https://doi.org/10.3847/1538-4365/aae9e1
Wolf ET, Haqq-Misra J, Toon OB. Evaluating climate sensitivity to $CO_2$ across Earth's history. *J Geophys Res (Atmos)* 2018;123(21):11861–11874; https://doi.org/10.1029/2018JD029262
Woodward HL, Rushby AJ, Mayne NJ. A novel metric for assessing climatological surface habitability. *Planet Sci J* 2025; 6;206–221; https://doi.org/10.3847/PSJ/adf3ab
Yang J, Cowan NB, Abbot DS. Stabilizing cloud feedback dramatically expands the habitable zone of tidally locked planets. *Astrophys J* 2013;771(2):L45–L50; https://doi.org/10.1088/2041-8205/771/2/L45
Yang J, Bouè G, Fabrycky DC, et al. Strong dependence of the inner edge of the habitable zone on planetary rotation rate. *Astrophys J* 2014;787:L2–L8; https://doi.org/10.1088/2041-8205/787/1/L2
Yang J, Zhang Y, Fu Z, et al. Cloud behaviour on tidally locked rocky planets from global high-resolution modelling. *Nat Astron* 2023;7:1070–1080; https://doi.org/10.1038/s41550-023-02015-8
Zhang Y, Yang J. How does background air pressure influence the inner edge of the habitable zone for tidally locked planets in a 3D view? *Astrophys J* 2020; 901(2);L36–L43; https://doi.org/10.3847/2041-8213/abb87f
Zechmeister M, Kürster M, Endl M. The M dwarf planet search programme at the ESO VLT + UVES. A search for terrestrial planets in the habitable zone of M dwarfs. *Astron Astrophys* 2009;505(2):859–871; https://doi.org/10.1051/0004-6361/200912479